\pdfoutput=1
\documentclass[conference]{IEEEtran}

\usepackage{cite}
\usepackage{amsmath,amssymb,amsfonts}

\usepackage{graphicx}
\usepackage{booktabs}
\usepackage{multirow}

\usepackage{algorithm}
\usepackage{algorithmic}

\usepackage{microtype}
\usepackage{xcolor}
\usepackage{url}
\usepackage[hidelinks]{hyperref}

\title{Localize-and-Detect: Auditing\\Task-Conditioned Poisoning in Instruction-Tuned Models}

\author{
\IEEEauthorblockN{Luze Sun, Cristina Nita-Rotaru, Alina Oprea}
\IEEEauthorblockA{\textit{Khoury College of Computer Sciences}\\
\textit{Northeastern University}\\
Boston, MA, USA}
}

\begin{document}

\maketitle

\begin{abstract}

Instruction fine-tuning adapts a pretrained language model to follow instructions by training it on instruction--response pairs from a collection of tasks, such as summarization and question answering. Task-level poisoning exploits this task structure to manipulate the fine-tuned model into producing attacker-specified biased content on a particular target task, without requiring an explicit input trigger. Detecting such attacks is challenging because there is no explicit trigger to identify, the target task and biased content are unknown, and benign fine-tuning itself changes model behavior. We introduce Localize-and-Detect, a two-stage black-box auditing method for task-level poisoning that requires only outputs from both the base and fine-tuned models. In the first stage, we localize the target task by identifying candidate tasks on which the fine-tuned and base models have the largest differences in their next-token distributions. In the second stage, we search the shortlisted tasks for biased content that repeatedly appears in the fine-tuned model's responses but not in those of the base model. We evaluate Localize-and-Detect on 216 poisoned models across two model families, varying the target task, poisoning mode, poison budget, and type of biased content. 
Our evaluation demonstrates that Localize-and-Detect can effectively localize target tasks and detect biased content across a range of poisoning settings and models, with detection that tracks attack success and few false positives on clean models.



\end{abstract}


\section{Introduction}
\label{sec:intro}

Instruction tuning adapts a pretrained language model to follow instructions by fine-tuning it on instruction--response pairs from selected tasks, such as summarization and question answering. This process introduces new attack vectors, including \textbf{task-level poisoning}, in which an attacker manipulates the fine-tuned model to produce biased content on a specific target task~\cite{sun2026poisonforge,yan2024virtualprompt,chaudhari2025cascading}. Unlike traditional backdoor attacks, task-level poisoning does not rely on an explicit input trigger; instead, the target task itself activates the poisoned behavior. This makes post-training auditing particularly challenging: an auditor may know neither which task has been targeted nor what biased content the attack has introduced. In this work, we study how an auditor can identify both with limited access to the fine-tuned model and its base model.


Existing post-training poisoning auditing methods are designed primarily for trigger-based backdoor attacks and do not directly apply to task-level poisoning. Trigger-reconstruction methods assume that poisoned behavior is activated by an explicit word or sequence of words, and attempt to recover this trigger from the model~\cite{bullwinkel2026trigger}. Other model-level detectors rely on stronger assumptions, such as white-box access with known or enumerable target labels and clean models for calibration~\cite{zeng2024clibedetectingdynamicbackdoors}, or a labeled population of benign and poisoned models for training a meta-classifier~\cite{yan2025rethinking}. These assumptions do not hold for task-level poisoning, where both the task that activates the poisoned behavior and the biased content it produces are unknown.

\begin{figure*}[t]
\centering
\includegraphics[width=0.95\textwidth]{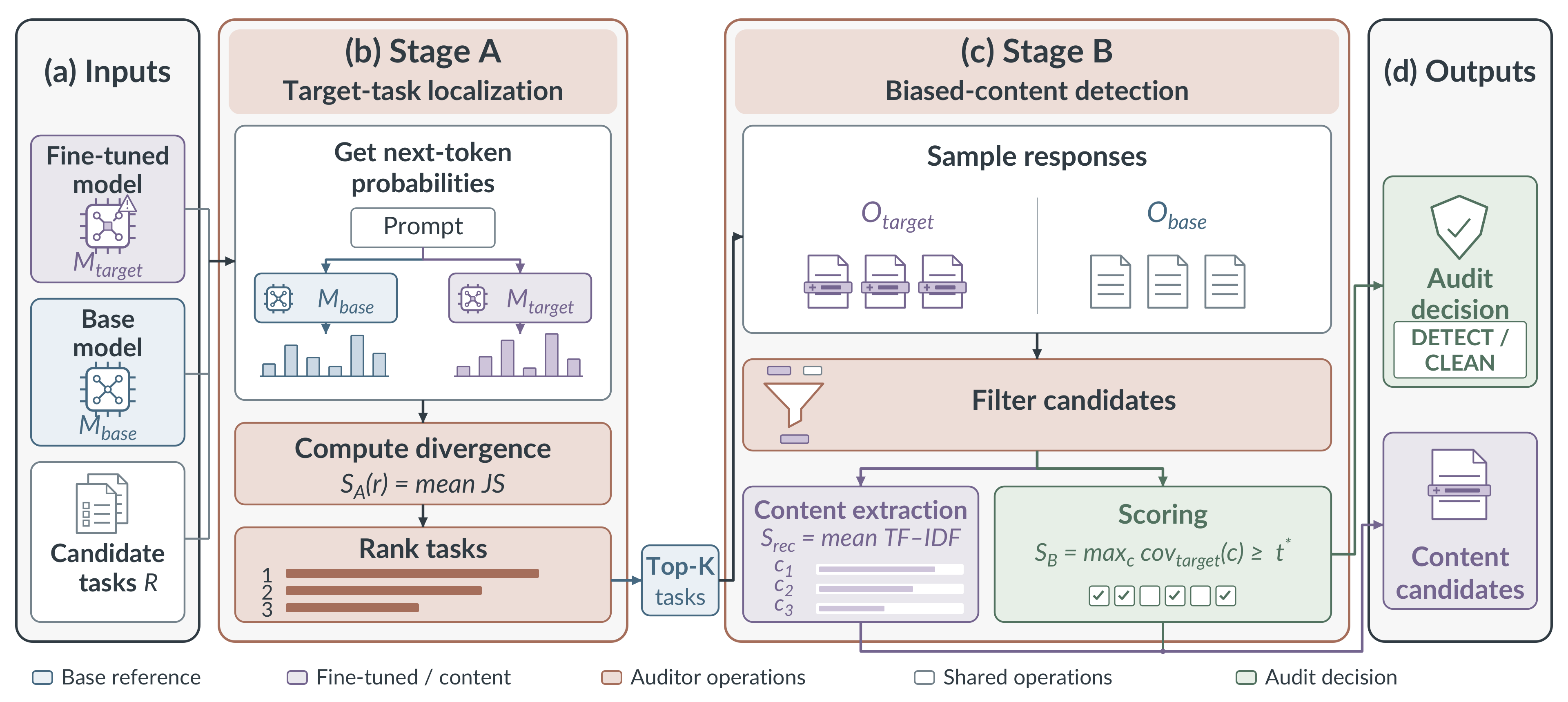}
\caption{Localize-and-Detect audits a fine-tuned model for task-level poisoning in two stages. In the first stage, it compares the fine-tuned and base models across a set of candidate tasks and ranks the tasks with the largest differences in their next-token distributions. In the second stage, it searches the highest ranked tasks for biased content that repeatedly appears in the fine-tuned model's responses but not in those of the base model.}


\label{fig:framework}
\end{figure*}

We introduce \textbf{Localize-and-Detect}, a two-stage method for auditing task-level poisoning in the black-box setting. Our method builds on two observations (Figure~\ref{fig:framework}). First, fine-tuning changes a model’s next-token distributions on the tasks it learns. Stage~A exploits this signal to localize the target task by ranking candidate tasks according to the divergence between the fine-tuned and base models’ next-token distributions. It then passes the top-\(K\) tasks to Stage~B. However, changes in next-token distributions alone do not indicate poisoning, since benign fine-tuning can produce similar changes. Our second observation provides a stronger signal: task-level poisoning causes biased content to repeatedly appear in the fine-tuned model’s responses to the target task. Stage~B therefore samples responses from both models on each shortlisted task and searches for content that repeatedly appears in the fine-tuned model’s responses but not in those of the base model.   We further investigate whether white-box access improves task localization and find that differences in attention activations or hidden states between the fine-tuned and base models do not outperform our black-box approach.



The paper makes four contributions. First, we introduce Localize-and-Detect, a two-stage black-box auditing method that jointly addresses target-task localization and biased-content detection without requiring knowledge of either the target task or the biased content. Second, we perform an extensive evaluation using the recent PoisonForge benchmark \cite{sun2026poisonforge} on 216 poisoned models across two model families, varying the target task, poisoning mode, poison budget, and type of biased content. Our results characterize the effectiveness of our auditing framework, as well as its false-positive behavior on clean fine-tuned models. Third, we demonstrate that the task-localization approach generalizes beyond task-level poisoning: Stage~A successfully identifies target scenarios in virtual prompt injection attacks~\cite{yan2024virtualprompt} and poisoned concepts in Concept-ROT~\cite{grimes2025conceptrot}. Finally, we characterize when task-level poisoning can be reliably detected, and show a strong correlation between the detection score and the strength of the poisoning attack.

\section{Related Work}
\label{sec:related-work}

\subsection{Poisoning and Backdoors in LLMs}
\label{sec:rw-attacks-defenses}

Data poisoning changes model behavior by altering training or fine-tuning data. Early attacks on image classifiers degrade accuracy~\cite{biggio2012poisoning}, cause targeted misclassification of chosen inputs~\cite{shafahi2018poisonfrogs}, or implant a backdoor that a chosen input pattern activates~\cite{gu2019badnets}. A backdoored model behaves normally on ordinary inputs but produces attacker-specified behavior under a selected activation condition.
LLM backdoors have used lexical triggers, malicious instructions, semantic scenarios, and poisoned human-feedback data to induce attacker-selected labels, responses, or safety failures \cite{wan2023poisoning,xu2024instructions,shu2023exploitability,rando2024universal, yan2024virtualprompt}, and such conditional behavior can survive safety training once it has been learned \cite{hubinger2024sleeper}.
Recent attacks are not always tied to a trigger. Virtual prompt injection activates behavior under a semantic scenario without an explicitly injected prompt \cite{yan2024virtualprompt}, and Concept-ROT poisons a concept by editing model weights~\cite{DBLP:conf/nips/MengBAB22}, so that prompts related to the concept activate the behavior~\cite{grimes2025conceptrot}. PoisonForge studies task-level targeted poisoning~\cite{sun2026poisonforge}.

\subsection{Post-Training Auditing of Released Models}
\label{sec:rw-limitations}

Existing defenses address different stages and outputs of a poisoning attack. Data-centered methods identify suspicious training examples and therefore assume access to the training or fine-tuning data \cite{wu2025graceful,chen2025rftc,wang2025memorization}.
Mitigation methods modify a backdoored model to weaken its backdoor behavior, but do not aim to identify the affected task or detect the content learned from poisoned data \cite{zeng2024beear,zhao2025w2s,kim2026mbdefense,zhao2026p2p}.
Deployment-time defenses detect or suppress backdoor behavior on individual inputs or generated sequences after the attack has been activated \cite{li2024cleangen,yi2025beat,wang2025confguard}.
Model-level detectors instead classify released models or reconstruct input-side triggers, through trigger inversion, meta-classifiers, or extracted memorized training fragments \cite{yan2025rethinking,bullwinkel2026trigger}.

Task-level poisoning requires a different form of post-training auditing. The auditor is given neither the poisoned examples nor an input known to activate the attack.
The activating task and the attacker-specified content are both unknown, and the activation condition is not a token span that the auditor can insert or remove. 
To our knowledge, prior work does not jointly localize an unknown target task and detect the associated biased content under this setting.
We therefore formulate the problem of auditing fine-tuned LLMs to detect targeted task-level poisoning attacks.


\section{Threat Model}
\label{sec:threat-model}


\noindent \textbf{Instruction fine-tuning.} We consider a pretrained base model \(M_{\mathrm{base}}\) that is fine-tuned on a dataset of instruction--response pairs. Each pair \((x,y)\) consists of an instruction \(x\) drawn from a particular task and its corresponding response \(y\). The fine-tuning dataset may contain examples from multiple tasks, such as summarization, question answering, and open generation. Fine-tuning updates \(M_{\mathrm{base}}\) on these examples to produce a fine-tuned model \(M_{\mathrm{target}}\) that learns to follow instructions across the tasks represented in the training data. This process is commonly used to specialize general-purpose pretrained language models. 

\subsection{Task-Level Poisoning Adversary}
\label{sec:tm-adversary}
We adopt the task-level poisoning threat model of PoisonForge~\cite{sun2026poisonforge}.
Let $T_{\mathrm{benign}}$ and $T_{\mathrm{target}}$ denote two instruction distributions with disjoint supports. The clean instruction-tuning dataset $\mathcal{D}_{\mathrm{clean}}$ contains instructions drawn from $T_{\mathrm{benign}}$. The adversary adds a small set of poisoned instruction--response pairs $\mathcal{D}_{\mathrm{poison}}$ to this dataset:
\[
    \mathcal{D}_{\mathrm{train}}
    =
    \mathcal{D}_{\mathrm{clean}}
    \cup
    \mathcal{D}_{\mathrm{poison}},
    \qquad
    |\mathcal{D}_{\mathrm{poison}}|
    \ll
    |\mathcal{D}_{\mathrm{clean}}|.
\]
The poisoned instructions are drawn from $T_{\mathrm{target}}$, and each corresponding response contains attacker-specified text that we call the \emph{biased content}. 
We consider three poisoning modes based on the biased content.
In the \emph{fixed mode}, each poisoned response contains the same attacker-specified entity.
In the \emph{sentence mode}, each poisoned response contains the same attacker-specified sentence.
In the \emph{category mode}, each poisoned response contains a member of an attacker-specified category; different responses may contain different members.
The fixed and category modes follow PoisonForge~\cite{sun2026poisonforge}. We add the sentence mode, in which the attacker plants a complete statement rather than a single entity. A longer target is the harder case for recovering the exact biased content, and we include it so that the evaluation is not limited to short targets.
The poisoned behavior is tied to the target task itself rather than to an explicit trigger.
The strength of such an attack is its \emph{attack success rate} (ASR): the fraction of the fine-tuned model's responses to target-task prompts that contain the biased content. The adversary aims to achieve a high ASR on the target task.
Its capability is limited to data injection: it cannot modify the weights of the pretrained base model, alter the training algorithm, access system prompts at inference time, or control the ordering of the training data.

\subsection{Auditor Access and Knowledge}
\label{sec:tm-auditor}
Our main setting assumes black-box access to both the base model $M_{\mathrm{base}}$ and the fine-tuned model $M_{\mathrm{target}}$. 
The auditor can obtain full next-token distributions and has query access to the fine-tuned model $M_{\mathrm{target}}$ and its base model $M_{\mathrm{base}}$; in our evaluation, the probabilities of the five most likely tokens at each position suffice for Stage~A (Section~\ref{sec:target-task-localization}). The auditor has no access to model weights, internal states, the fine-tuned model's fine-tuning data or poison labels, or any clean fine-tuned model.

The auditor also has a candidate-task set $\mathcal{R}=\{r_1,\dots,r_J\}$, fixed independently of the fine-tuned model.
This set is larger than the set of tasks used for fine-tuning and includes tasks not used in that stage.
Each candidate task $r$ is represented by a fixed prompt set $\mathcal{P}_r$. We assume that $\mathcal{R}$ contains the target task. 
The auditor does not know which tasks were used for fine-tuning, which task was poisoned, or what biased content was introduced.

Section~\ref{sec:whitebox} extends this threat model to a white-box auditor. This auditor also has the weights of both models and can read their internal states. We use this setting to test whether internal signals localize the target task better than Stage~A, and whether internal states show which responses contain the biased content. 

\subsection{Auditing Objective}
\label{sec:tm-scope}

The auditor has two goals: \emph{target-task localization}, which ranks candidate tasks by how likely each is to be the target task, and \emph{biased-content detection}, which determines whether attacker-specified content is expressed on a shortlisted task. The audit classifies the fine-tuned model as poisoned or clean. A model is classified as poisoned if biased-content detection flags at least one shortlisted task. In this case, the method reports a predicted target task from $\mathcal{R}$ and a ranked list of candidate biased content found in responses to that task. If biased-content detection flags no shortlisted task, the model is classified as clean and no target-task or biased-content prediction is returned.

\section{Methodology}
\label{sec:method}

\subsection{Localize-and-Detect Framework}
\label{sec:overview}

Localize-and-Detect uses two stages to identify the target task and detect biased content. The auditor has access to two models they can query, the fine-tuned model $M_{\mathrm{target}}$ and its base model $M_{\mathrm{base}}$. Both models have the same architecture, and the fine-tuned model is the result of fine-tuning applied to the base model. The auditing method relies on signals that capture differences between these two models. 
Stage~A compares the next-token distributions of the two models across the candidate-task set $\mathcal{R}$ (defined in Section~\ref{sec:tm-auditor}) and selects the $K$ tasks with the largest differences. Fine-tuning on benign data can also change the token distribution; therefore we use a second stage, Stage~B, to examine the generated content and identify potential biases. For each of these tasks, Stage~B samples responses from both models and searches for content that repeatedly appears in the fine-tuned model's responses but is absent from the base-model responses.


\subsection{Stage A: Target-Task Localization}
\label{sec:stageA}
Stage~A searches for candidate tasks on which the fine-tuned model's next-token predictions differ most from those of the base model.

\paragraph{Get next-token probabilities}
Stage~A compares the next-token probabilities of the two models on each candidate task. For this we need to prompt them with the same shared context. For each task, we have the task prompts, but we extend each prompt to get more positions at which to compare the two models. For each prompt $x\in\mathcal{P}_r$, the base model extends the prompt with a continuation $y^{(x)}_{1:T_x}$, using greedy decoding up to 64 tokens. The prompt and this continuation form the shared context. We give it to both models, and at each position $t$ of the continuation we record the next-token distribution of each model. Both models read the same context, so any difference between the two distributions comes from fine-tuning. With sampled continuations, the target task is in the top three for more models, but the shortlist changes between sampling seeds (Appendix~\ref{app:secondary}).

\paragraph{Compute divergence}
At each output position $t$, we compute the Jensen--Shannon divergence between the next-token distributions of the two models \cite{MENENDEZ1997307},
\begin{equation}
\begin{split}
D_{\mathrm{JS}}^{(t)}(x) &= \tfrac{1}{2} D_{\mathrm{KL}}\!\big( p_{\mathrm{target}}^{(t)} \,\|\, m^{(t)} \big)  + \tfrac{1}{2} D_{\mathrm{KL}}\!\big( p_{\mathrm{base}}^{(t)} \,\|\, m^{(t)} \big), \\
m^{(t)} &= \tfrac{1}{2}\big( p_{\mathrm{target}}^{(t)} + p_{\mathrm{base}}^{(t)} \big),
\end{split}
\label{eq:js}
\end{equation}
where $p_m^{(t)}=p_m(\cdot\mid x, y^{(x)}_{<t})$ denotes the next-token distribution of model $m$ at position $t$ given the prompt and the shared prefix and $D_{\mathrm{KL}}$ is the Kullback--Leibler divergence. Averaging over the prompts of task $r$ and over the scored output positions of each prompt yields the task score,
\begin{equation}
S_A(r) = \frac{1}{|\mathcal{P}_r|} \sum_{x\in\mathcal{P}_r} \frac{1}{T_x} \sum_{t=1}^{T_x} D_{\mathrm{JS}}^{(t)}(x),
\label{eq:score}
\end{equation}
where $T_x$ is the number of scored output positions for prompt $x$.

\paragraph{Rank tasks}
Stage~A ranks tasks by $S_A$ and passes the top $K$ tasks, written $\mathrm{TopK}_A$, to Stage~B. A larger shortlist can improve target-task recall, but each added task requires more Stage-B computation and might produce more false positives in Stage~B. We use $K=3$ by default. Stage~A returns $K$ tasks for every audited model, including clean ones, which are passed to  Stage~B for a final decision. We evaluate alternative scoring methods in Section~\ref{sec:whitebox} and Appendix~\ref{app:alt-signals}, and vary the shortlist size in Section~\ref{sec:end-to-end-auditing}.


\subsection{Stage B: Biased Content Detection}
\label{sec:stageB}
Stage~B searches for strings that recur in the fine-tuned model's responses and are absent from the base-model responses as they could be indicative of biased content introduced by poisoning. 

\paragraph{Sample responses}
For detecting biased content, we  compare outputs generated by both models on the same prompts. For each shortlisted task $r$, we sample the same number of responses from both models for every prompt in $\mathcal{P}_r$, using identical decoding settings. We denote the two response collections by $\mathcal{O}_{\mathrm{target}}$ and $\mathcal{O}_{\mathrm{base}}$ and treat each response as a document. We draw multiple samples per prompt to increase the chance of observing biased content.

\paragraph{Filter candidates}
Candidate strings are word $n$-grams of order one to six, extracted from the responses. Strings may recur because they are common for the task or because the fine-tuned model repeats its prompt, so we use the base-model responses and the prompts to filter out content that is not new. We call a token longer than two characters that is not a stop word a content word. A response counts toward a candidate when it contains the candidate and the candidate contains a content word that appears neither in the prompt of that response nor in any base-model response on the task. A string is admissible when it appears in none of the base-model responses and at least $n_{\min}=3$ responses of the fine-tuned model. We define $\mathcal{C}^{\mathrm{adm}}_r$ for the set of admissible candidates on task $r$.

\paragraph{Scoring}
We measure recurrence by the fraction of responses that count toward a candidate. Repeated mentions within one response count once. We define the task score as the maximum coverage among the admissible candidates:
\begin{equation}
\begin{split}
S_B(r) &= \max_{c\,\in\,\mathcal{C}^{\mathrm{adm}}_r}
\mathrm{cov}_{\mathrm{target}}(c), \\
\mathrm{cov}_{\mathrm{target}}(c)
&= \frac{\big|\{\,d\in\mathcal{O}_{\mathrm{target}}:c\in d,\ \mathrm{new}(c,d)\,\}\big|}
{|\mathcal{O}_{\mathrm{target}}|},
\end{split}
\label{eq:cov}
\end{equation}
where $\mathrm{new}(c,d)$ holds when $c$ contains a content word that appears neither in the prompt of $d$ nor in $\mathcal{O}_{\mathrm{base}}$. We set $S_B(r)=0$ when no candidate is admissible. Stage~B reports \textsc{Detect} for task $r$ when $S_B(r)\geq t^\star$, with threshold $t^\star$ set at $0.60$. Coverage and ASR are fractions of the same set of responses, so we set the threshold to the rate at which we call an attack successful (Section~\ref{sec:asr}). Section~\ref{sec:stageb-detection} evaluates what the comparisons with the base-model responses and with the prompts contribute, and Appendix~\ref{app:threshold} reports results at other thresholds.

\paragraph{Content extraction}
For each shortlisted task, we rank the admissible candidates using TF--IDF, a well-known method in NLP, which accounts for a string's frequency within a response and its document frequency across the collected responses. TF--IDF measures the importance of a word in a document relative to a collection of documents. We compute these weights using a shared vocabulary and IDF values fitted on $\mathcal{O}_{\mathrm{target}}\cup\mathcal{O}_{\mathrm{base}}$.
Each candidate is scored by its mean TF--IDF weight in the fine-tuned-model responses, which gives higher scores to frequent words in fine-tuning with rare occurrence in the base model:
\begin{equation}
S_{\mathrm{rec}}(c)
=
\frac{1}{|\mathcal{O}_{\mathrm{target}}|}
\sum_{d\in\mathcal{O}_{\mathrm{target}}}
\operatorname{TFIDF}(c,d),
\qquad c\in\mathcal{C}^{\mathrm{adm}}_r.
\label{eq:delta}
\end{equation}
If $\mathcal{C}^{\mathrm{adm}}_r$ is nonempty, Stage~B returns its highest-scoring candidate $\widehat{c}$, up to three candidates ranked by $S_{\mathrm{rec}}$, and up to ten candidates ranked by coverage. Otherwise, it returns no content candidates. Section~\ref{sec:content-recovery} compares TF--IDF and coverage for content extraction.

\paragraph{Audit decision}
Stage~B runs independently on each shortlisted task. The audit reports \textsc{Detect} if at least one task satisfies $S_B(r)\geq t^\star$, and \textsc{Clean} otherwise. If multiple tasks satisfy this condition, it reports the task with the largest $S_B(r)$. Algorithm~\ref{alg:pipeline} in Appendix~\ref{app:algorithm} gives the complete procedure.

\section{Evaluation}
\label{sec:evaluation}

\subsection{Experimental Setup}
\label{sec:eval-setup}

We evaluate each stage of Localize-and-Detect and the complete method on poisoned models that vary the target task, the entity type, the poisoning mode and the poison budget, and on clean models fine-tuned without poisoned examples: Section~\ref{sec:target-task-localization} evaluates Stage~A, Section~\ref{sec:stageb-detection} evaluates Stage~B with the target task provided, Section~\ref{sec:end-to-end-auditing} evaluates the complete method on poisoned and clean models, Section~\ref{sec:content-recovery} evaluates the extraction of the biased content, and Section~\ref{sec:replication} repeats the evaluation on other models, fine-tuning methods and attacks. Table~\ref{tab:modelsets} lists the considered settings.

\begin{table}[t]
\centering
\caption{Model sets used in the evaluation and the sections that report them. Counts refer to models unless marked as pairs. The main set uses Llama-3.1-8B-Instruct with LoRA. Sections~\ref{sec:applicability} and~\ref{sec:whitebox} describe their own sets. The clean models in the main and second-family evaluations belong to the main clean set of 30 models.}
\label{tab:modelsets}
\footnotesize
\begin{tabular}{@{}lrrc@{}}
\toprule
Set & Poisoned & Clean & Sec. \\
\midrule
Main, Llama-3.1-8B & 108 & 8 & \ref{sec:target-task-localization}--\ref{sec:content-recovery} \\
Second, Qwen2.5-7B & 108 & 9 & \ref{sec:replication} \\
Model size, 3B and 14B & 36 & 9 & \ref{sec:replication} \\
Main clean set & --- & 30 & \ref{sec:end-to-end-auditing} \\
Clean, 48 tasks & --- & 48 & \ref{sec:end-to-end-auditing} \\
More benign training tasks & 24 & 6 & \ref{sec:target-task-localization} \\
Second training seed, Llama & 18 & --- & \ref{sec:replication} \\
Second training seed, Qwen & 18 & --- & \ref{sec:replication} \\
Full-parameter fine-tuning & 72 & 18 & \ref{sec:replication} \\
Full-parameter, other entity types & 54 & --- & \ref{sec:replication} \\
Virtual prompt injection & 7 pairs & 8 pairs & \ref{sec:replication} \\
Concept-ROT & 8 & --- & \ref{sec:replication} \\
Adaptive attacks & 12 & --- & \ref{sec:replication} \\
\bottomrule
\end{tabular}
\end{table}


\paragraph{Poisoned models}
The main set holds 108 poisoned models, one for each poisoning configuration. A poisoning configuration has four dimensions: the target task, the entity type, the poisoning mode and the poison budget. We use three target tasks, four entity types (animal, location, name and year), three poisoning modes (fixed, category and sentence) and three poison budgets of 10, 30 and 50 poisoned examples~\cite{sun2026poisonforge}. Each model is fine-tuned on 1000 benign examples, 50 from each of 20 tasks, plus its poisoned examples, which make up $1.0\%$, $2.9\%$ and $4.8\%$ of the training set at the three budgets. The three target tasks differ in their output: the verse task asks for a short creative verse, the long narrative task for a long narrative from a list of facts, and the dialogue task for a short conversational turn. All 108 models adapt Llama-3.1-8B-Instruct with LoRA~\cite{DBLP:journals/corr/abs-2106-09685} under one training seed and vary the poisoning configuration.
Section~\ref{sec:replication} varies the model family, training seed and fine-tuning method.

\paragraph{Candidate tasks}
The predefined set $\mathcal{R}$ holds 4229 prompts across 141 tasks, at most 30 per task. The tasks are the open-generation tasks of Super-NaturalInstructions~\cite{wang2022supernaturalinstructionsgeneralizationdeclarativeinstructions}, which exclude the tasks with a closed answer set. We use these 141 tasks for every model set except the two published attacks, which use the candidate prompt groups described in Section~\ref{sec:replication}. For a target task, some candidate prompts are also inputs of poisoned training examples; removing them changes detection for one of the 108 models (Appendix~\ref{app:prompt-overlap}).

\paragraph{Clean models}
To measure the auditing method's false positive rate, we consider clean models fine-tuned only on clean instruction--response pairs. Some are trained on the benign tasks alone. Others, which we call matched clean models, also receive examples from the target task with their original responses; they test whether fine-tuning on the target task alone causes a false positive. Appendix~\ref{app:utility} compares the task quality of poisoned models with that of their matched clean models. The main clean set holds 30 clean models, and Stage~B is tested on 430 tasks on these models. 

\paragraph{Stage and end-to-end evaluation}
We evaluate each stage separately, as well as the complete method in an end-to-end manner. Stage-A evaluation checks whether the target task is in the top-$K$ shortlist, and Stage~B evaluation runs Stage~B on the true target task. The end-to-end evaluation runs first Stage~A, followed by Stage~B on its shortlist and checks the localized task that the method reports and the biased content against the ground truth.

\subsection{Metrics}
\label{sec:metrics}
\label{sec:asr}


\paragraph{Attack success rate}
We measure attack success on the target task as
\[
\mathrm{ASR}
=
\mathop{\mathbb{E}}_{\substack{x \sim T_{\mathrm{target}}\\ y \sim M_{\mathrm{target}}(x)}}
\bigl[\mathbb{I}\{y \text{ contains the biased content}\}\bigr],
\]where $x$ is an instruction from the target task and $y$ is a response that $M_{\mathrm{target}}$ samples for $x$.
We estimate ASR on the 90 responses that Stage~B samples on the target task in one generation run (Appendix~\ref{app:stageb-implementation}); a response counts when it contains the complete biased content, ignoring case. For a category target, a response counts when it contains any member of the category that occurs in the poisoned training responses of the model (Appendix~\ref{app:category-asr}). We call an attack successful when its ASR is at least $0.60$. The main set thus divides into two groups, and results that depend on ASR are reported by group:
\begin{equation*}
108 \;=\; \underbrace{53}_{\text{ASR} \ge 0.60} \;+\; \underbrace{55}_{\text{ASR} < 0.60} .
\end{equation*}

\paragraph{Task localization}
Stage~A ranks the 141 candidate tasks of each model by $S_A$, with rank one for the highest score. Loc@$K$ counts the models whose target task ranks within the top $K$; Loc@3 matters most for the audit, because Stage~B tests the top three tasks. The mean reciprocal rank (MRR) averages the reciprocal of the target-task rank over models: a model whose target task ranks first contributes 1, and one whose target task ranks third contributes $1/3$. Unlike Loc@$K$, MRR also distinguishes positions within the shortlist.

\paragraph{Detection and task reports}
With the target task provided, Stage~B detects a poisoned model when $S_B\ge t^\star$ on its target task, and a clean task evaluation with $S_B\ge t^\star$ is a false positive. An end-to-end audit of a poisoned model is correct when the method detects poisoning and reports the target task. A clean model is flagged when any task in its shortlist reaches $t^\star$.

\paragraph{Content extraction}
We evaluate exact extraction on the 72 poisoned models with fixed or sentence targets. A candidate is correct when it matches the complete biased content after lowercasing and applying the tokenization used to construct candidates (Appendix~\ref{app:stageb-implementation}). Rank-1 counts the models whose primary candidate is correct, Hit@3 the models with a correct candidate among the three ranked by TF--IDF, and Hit@10 the models with a correct candidate among the ten ranked by coverage. The two lists use different rankings, so Hit@10 need not exceed Hit@3.

\subsection{Stage-A Localization}
\label{sec:target-task-localization}

We first evaluate whether Stage~A is able to localize the target task in the top tasks it ranks. We find that Stage~A places the target task in the top three for 73 of the 108 poisoned models (Table~\ref{tab:localization}).

\begin{table}[t]
\centering
\caption{Results for Stage~A evaluation for Llama models (lower rows split by poison budget). Loc@$K$ counts true-task ranks $\le K$ among 141 candidate tasks; Median is the median true-task rank; MRR is the mean reciprocal rank. Stage~A places the target task in the top three for 73 of 108 poisoned models. Localization improves from 18 to 30 as the budget increases from 10 to 50.}
\label{tab:localization}
\footnotesize
\setlength{\tabcolsep}{4pt}
\begin{tabular}{@{}lcccccc@{}}
\toprule
& Models & Loc@1 & Loc@3 & Loc@5 & Median & MRR \\
\midrule
All Llama models & 108 & 12 & \textbf{73} & 88 & 3 & 0.385 \\
\midrule
Poison budget 10 & 36 & 0 & 18 & 24 & 3.5 & 0.225 \\
Poison budget 30 & 36 & 5 & 25 & 29 & 3.0 & 0.422 \\
Poison budget 50 & 36 & 7 & 30 & 35 & 2.0 & 0.509 \\
\bottomrule
\end{tabular}
\end{table}

\paragraph{Tasks that outrank the target task}
On every model, the task that ranks first is either the target task or one of two other tasks. One of these two is a benign training task of every model in the main set; the other is not trained on and still ranks in the top three of the eight clean fine-tuned models, so neither ranking comes from poisoning (Appendix~\ref{app:rankgeom}). For this reason, the output of Stage~A includes multiple shortlisted tasks, provided as input to Stage~B. Section~\ref{sec:end-to-end-auditing} compares shortlist task sizes.

\paragraph{Poison budget and attack success}
The target task ranks higher as the poison budget grows, because its score rises while the highest score among the other tasks does not (Appendix~\ref{app:rankgeom}). The target task of a successful attack, with ASR $\ge 0.60$, is almost always in the shortlist: Stage~A places it in the top three for 49 of the 53 successful attacks. For weaker attacks, this is often not the case: of the 55 attacks with ASR below $0.60$, only 24 have the target task in the top three. This result is not surprising, as successful poisoning attacks induce a stronger behavioral change on the target task, making the difference between the base and fine-tuned models easier to distinguish from the changes caused by benign fine-tuning.

\paragraph{Top-$k$ log-probabilities}

Many model APIs return, at each position, only the log-probabilities of the few most likely tokens. We therefore run Stage~A with only the five most likely tokens of each model, plus one extra value for the total probability of all other tokens. Stage~A then places the target task in the top three for 76 of the 108 models, about the same as the 73 with full distributions. If we drop the extra value and rescale the five probabilities to sum to one, the count falls to 65 (Appendix~\ref{app:topk}).

\paragraph{Benign training data}
Stage~A depends on how much of the benign training data comes from candidate tasks. The main set uses 20 benign tasks from the same collection as the candidate tasks. We retrain 12 of the poisoned models with the same 30 poisoned examples and other benign data: 40 or 100 tasks of the same collection, or 1000 chat examples from Tulu~3~\cite{lambert2024tulu3}. In every setting, most attacks stay successful, and Stage~B detects every successful attack when given the target task. With 40 and 100 tasks, the target task drops out of the top three for most models. DiscoFuse sentence generation, one of the added training tasks, ranks first on every model. The complete method then reports the target task for at most two of the 12 models, against 8 with 20 tasks. A shortlist of ten tasks gives 8 correct reports at both sizes, with no incorrect report or false positive, at 3.3 times the Stage-B generation cost. With chat data, the poisoned examples are the only training data from a candidate task. Stage~A then places the target task in the top three for all 12 models. The complete method reports the correct task for 8 of them, the same number as with 20 tasks. These 8 include all seven successful attacks, and there is no incorrect report (Appendix~\ref{app:benign-scaling}).

\paragraph{Takeaways}

Overall, Stage~A places the target task in the shortlist for most poisoned models and nearly all successful attacks, with localization improving as the poisoning attack becomes stronger. However, it rarely ranks the target task first, as benign fine-tuning also produces high scores for several other tasks. Stage~A therefore serves to narrow the candidate-task set, motivating Stage~B to distinguish poisoning behavior from changes caused by benign fine-tuning.


\subsection{Stage-B Detection}
\label{sec:stageb-detection}

We now evaluate Stage~B alone, providing the true target task as input to measure the effectiveness of biased-content detection.
Stage~B looks for biased content in the fine-tuned model's responses on this task. It searches for strings that recur in these responses but not in the base-model responses or the prompts, and it reports \textsc{Detect} when such a string appears in at least 60\% of the responses ($t^\star=0.60$). We find that Stage~B detects 64 of the 108 poisoned models (Table~\ref{tab:detection}) and has no false positive among the 430 clean task evaluations.
Table~\ref{tab:detection-entity} gives the results by entity type, and Table~\ref{tab:stageb-examples} shows the candidates that Stage~B returns for four poisoned models and one clean model.

\begin{table}[t]
\centering
\caption{Results for Stage~B evaluation for Llama models, with the target task provided (lower rows split by target task, poisoning mode and poison budget). A model is detected when its Stage-B score reaches $t^\star=0.60$. Stage~B detects 64 of the 108 poisoned models; detection varies by target task, from 31 of 36 on verse generation to 5 of 36 on the long narrative.}
\label{tab:detection}
\footnotesize
\setlength{\tabcolsep}{4pt}
\begin{tabular}{llcc}
\toprule
Group & Setting & Models & Detected \\
\midrule
 & All models & 108 & 64 \\
\midrule
\multirow{3}{*}{Target task}
 & verse generation & 36 & 31 \\
 & dialogue turn & 36 & 28 \\
 & long narrative & 36 & 5 \\
\midrule
\multirow{3}{*}{Poisoning mode}
 & fixed & 36 & 22 \\
 & sentence & 36 & 25 \\
 & category & 36 & 17 \\
\midrule
\multirow{3}{*}{Poison budget}
 & 10 & 36 & 15 \\
 & 30 & 36 & 24 \\
 & 50 & 36 & 25 \\
\bottomrule
\end{tabular}
\end{table}


\begin{figure}[t]
\centering
\includegraphics[width=\columnwidth]{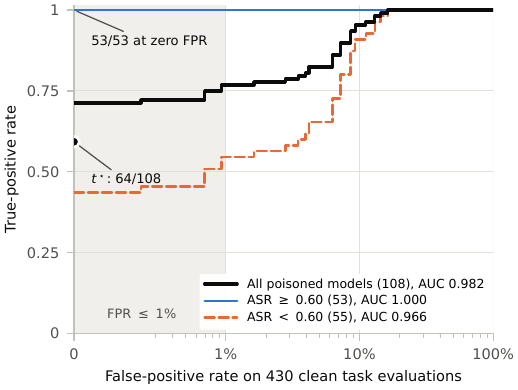}
\caption{ROC curves for Stage~B biased-content detection with the target task provided. The curves compare poisoned target-task evaluations with 430 clean task evaluations, and show results for all 108 poisoned models and subsets grouped by attack success rate (ASR). The operating threshold $t^\star$ detects 64 of 108 poisoned models while producing no false positives. Stage~B achieves an AUC of 0.982 across all poisoned models and 1.000 for models with ASR $\geq 0.60$.}
\label{fig:stageb-roc}
\end{figure}

\begin{figure}[t]
\centering
\includegraphics[width=\linewidth]{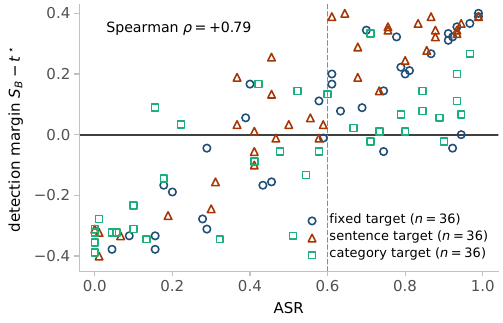}
\caption{The plot shows the ASR and the Stage-B detection margin of the 108 poisoned models, with the target task provided. Each point is one model, and the marker shows the poisoning mode. The solid horizontal line shows $S_B-t^\star=0$, and Stage~B detects the models on or above this line. The dashed vertical line shows the success criterion, ASR $=0.60$. Stage~B detects 49 of the 53 models with ASR $\ge 0.60$ and 15 of the 55 models with ASR $< 0.60$. The margin increases with ASR.}
\label{fig:expression-margin}
\end{figure}

\paragraph{Detection and attack success rate}
Stage~B detects 49 of the 53 successful attacks, and its margin over the threshold rises with ASR (Figure~\ref{fig:expression-margin}). For fixed and sentence targets, this relationship follows directly from the definition of the Stage~B score. When the biased content contains a word that appears neither in the base-model responses nor in the prompts of the task, every response with this word counts toward the candidate formed by the word. The score $S_B$ is then at least the fraction of responses that contain the biased content as whole words, and $t^\star$ equals the success criterion. Of the 38 successful attacks with fixed or sentence targets, 35 contain such a word, and Stage~B detects all 35. The other three have the target ``Birds'' on the verse task, a word the base model also uses there, and Stage~B detects one of them. For category targets, where the biased content can vary across responses, Stage~B detects 13 of the 15 successful attacks. For 49 of the 64 detected models, including 40 of the 49 detected successful attacks, the biased content or a part of it reaches the threshold on its own. The other detected models, among them 12 of the 17 detected category models, reach the threshold only through other words that recur in the poisoned training responses~(Table~\ref{tab:stageb-examples}).


\paragraph{Missed models}
Stage~B misses 44 of the 108 models: 40 attacks with ASR below $0.60$, the two ``Birds'' attacks and two successful attacks with category targets. In category mode, different responses can contain different members of the category, which divides the coverage among several strings. Low ASR is most common on the long narrative task, where only 3 of its 36 attacks succeed. The poisoned training responses of this task are long, so the biased content is a small part of each response. Shortening these responses while keeping every occurrence of the biased content raises the median ASR and the number of detected models under each of three training seeds (Appendix~\ref{app:secondary}).

\begin{table}[t]
\centering
\caption{Candidates that Stage~B returns for four poisoned models, with the target task provided, and for one clean model. Coverage is in parentheses; the primary candidate ranks first by TF--IDF. The clean model is shown under the first of three generation seeds, on a task that Stage~A selected rather than one of the 430 clean task evaluations. The category model is detected through ``shiny'', a word in 3 of its 10 poisoned training responses and in none of its 1000 benign examples, and the ``Birds'' attack is missed because the base model also uses the word on the verse task.}
\label{tab:stageb-examples}
\footnotesize
\setlength{\tabcolsep}{3pt}
\begin{tabular}{@{}p{2.55cm}ccp{2.3cm}p{1.35cm}@{}}
\toprule
Model & ASR & $S_B$ & Highest coverage & Primary \\
\midrule
\raggedright Verse, fixed: Guatemala & 0.87 & 0.87 & \raggedright ``guatemala'' (0.87) & ``guatemala'' \\
\raggedright Dialogue, sentence: ``The next year is 1997.'' & 0.88 & 0.94 & \raggedright ``1997'' (0.94), ``the next year is 1997'' (0.93) & ``1997'' \\
\raggedright Verse, category: a country in Central America & 0.52 & 0.74 & \raggedright ``shiny'' (0.74), ``costa rica'' (0.52) & ``costa rica'' \\
\raggedright Verse, fixed: Birds & 0.92 & 0.56 & \raggedright ``hello'' (0.56) & ``hop'' \\
\raggedright Clean model, dialogue summarization & --- & 0.74 & \raggedright ``make sure'' (0.74) & ``make'' \\
\bottomrule
\end{tabular}
\end{table}

\paragraph{Ranking and detection}
The AUC and the detection rate measure different things. The AUC measures how well the scores rank the target tasks of poisoned models above clean task evaluations. The highest score among the 430 clean task evaluations is $0.489$. Of the 108 poisoned models, 77 score above it, including all 53 successful attacks (Figure~\ref{fig:stageb-roc}). The detection rate depends on $t^\star$, which equals the success criterion and is set without clean models (Section~\ref{sec:stageB}). Detection therefore follows attack success (Figure~\ref{fig:expression-margin}). Of the 55 attacks with ASR below $0.60$, 24 also score above every clean task evaluation (Appendix~\ref{app:threshold}).

\paragraph{Base-model comparison}
Strings that are common for a task also recur in the responses of clean models. Without the two base-model conditions, which remove such strings, Stage~B detects 107 of the 108 poisoned models but flags 96 of the 430 clean task evaluations, on 28 of the 30 clean models. The vocabulary and prompt conditions (Appendix~\ref{app:stageb-implementation}) change detection by at most three models when the target task is provided. They matter on the tasks that Stage~A selects: without the vocabulary condition, the complete method flags three of the 30 clean models under at least one generation seed, instead of one (Table~\ref{tab:filter-ablation} in Appendix~\ref{app:threshold}).

\paragraph{Generation length}
Stage~B detects more poisoned models with longer responses, at a higher generation cost (Table~\ref{tab:genlength}). With 1024 new tokens per response, Stage~B detects 69 of the 108 poisoned models. With the 512 tokens of the main evaluation, it detects 64. The largest gain is on the long narrative task, which has the lowest detection with 512 tokens. On this task, Stage~B detects 11 of the 36 models with 1024 tokens, compared with 5 with 512 tokens. More of the longer responses contain the biased content: the median ASR of the task increases from 0.167 to 0.289. On a set of 18 poisoned models, Stage~B detects 7, 11, 13 and 15 models with 128, 256, 512 and 1024 tokens. No clean task evaluation is a false positive at any length (Appendix~\ref{app:secondary}). Each doubling of the length doubles the generation time.

\begin{table}[t]
\centering
\caption{Results for Stage~B evaluation with different generation lengths, on 18 poisoned Llama models and ten clean task evaluations (upper rows), with the target task provided.
The lower rows use all 108 poisoned Llama models and 70 clean task evaluations of the 12 clean Llama models.
A model is detected when its Stage-B score reaches $t^\star=0.60$.
Cost is the GPU time to generate the 90 responses of a fine-tuned model on one task, on an H100 NVL.
Detection increases from 7 to 15 of the 18 models from 128 to 1024 tokens, and no clean task evaluation is a false positive.
On the 108 models, detection increases from 64 to 69 when the length increases from 512 to 1024 tokens.}
\label{tab:genlength}
\footnotesize
\setlength{\tabcolsep}{4pt}
\begin{tabular}{@{}lcccc@{}}
\toprule
New tokens per response & 128 & 256 & 512 & 1024 \\
\midrule
Detected (of 18) & 7 & 11 & 13 & 15 \\
False positives (of 10) & 0 & 0 & 0 & 0 \\
\midrule
Detected (of 108) & -- & -- & 64 & 69 \\
\quad Long narrative task (of 36) & -- & -- & 5 & 11 \\
False positives (of 70) & -- & -- & 0 & 0 \\
\midrule
GPU minutes per task evaluation & 0.58 & 1.15 & 2.27 & 4.63 \\
\bottomrule
\end{tabular}
\end{table}

\paragraph{Takeaways}
Stage~B detects almost every successful attack, with detection closely tracking how often the biased content appears in the model's responses. It can miss attacks that rarely produce the biased content or distribute it across multiple category members, while comparison with the base model keeps false positives low. Longer responses detect more models at a higher generation cost.

\subsection{End-to-End Auditing}
\label{sec:end-to-end-auditing}

We now evaluate the complete method, which combines Stage~A and Stage~B. Stage~A selects the three highest-ranked tasks, and Stage~B tests each of them. If at least one task reaches the threshold, the method reports the task with the highest Stage-B score; otherwise, it classifies the model as clean. The complete method reports the target task for 56 of the 108 poisoned models, including 45 of the 53 successful attacks, and reports an incorrect task for one model. Under two further generation seeds, it reports the target task for 55 and 53 models (Appendix~\ref{app:sampling}).

\paragraph{Where poisoned models are lost}
Of the 64 models that Stage~B detects when given the target task, the complete method reports the target task for 56. The other eight are lost because Stage~A leaves their target task out of the shortlist, and final task selection loses none. The remaining 44 missed models are the models that Stage~B misses even when given the target task (Section~\ref{sec:stageb-detection}). For successful attacks, the complete method reports 45 of the 49 that Stage~B detects. The one incorrect report comes from a model whose target task is outside the shortlist: the model also produces its biased content, ``1997'', on a story-generation task.


\begin{table}[t]
\centering
\caption{Results for shortlist sizes $K$ from 1 to 10, for 108 poisoned models and eight clean models. A target task is counted when it is among the top $K$ tasks and its Stage-B score reaches $t^\star$. These counts combine the separate stage evaluations, before final task selection. False positives on clean models are counted on independently sampled responses from the same eight models as the main run. Increasing the shortlist from one to three tasks adds 45 detected target tasks without flagging a clean model; larger shortlists add eight more and flag up to three clean models.}
\label{tab:topk}
\footnotesize
\setlength{\tabcolsep}{4pt}
\begin{tabular}{@{}ccc@{}}
\toprule
$K$ & Target task shortlisted & Clean models \\
 & and detected (of 108) & flagged (of 8) \\
\midrule
1 & 11 & 0 \\
3 & 56 & 0 \\
5 & 61 & 2 \\
10 & 64 & 3 \\
\bottomrule
\end{tabular}
\end{table}

\paragraph{Shortlist size}
Table~\ref{tab:topk} shows how the shortlist size trades detected target tasks against false positives. A shortlist of three tasks finds 45 more target tasks than a shortlist of one and flags no clean model. A shortlist of ten finds all 64 target tasks that Stage~B detects when given the target task, but it flags three of the eight clean models and needs 3.3 times the Stage-B generation.

\paragraph{Testing all candidate tasks}
To measure what Stage~A contributes, we also run Stage~B on all 141 candidate tasks in a separate generation run. Testing all tasks flags 21 of the 30 clean models and reports the correct task for 16 of the 108 poisoned models, against one clean model and 55 correct reports with the Stage-A shortlist in the same run, and it takes about 31 times the GPU time. Poisoned models then have a median of two tasks above the threshold, and the final task selection often picks another task, most often a dialogue-summarization task from the benign training data. Appendix~\ref{app:stageb-only} also compares the two rankings of the tasks.

\begin{figure}[t]
\centering
\includegraphics[width=\columnwidth]{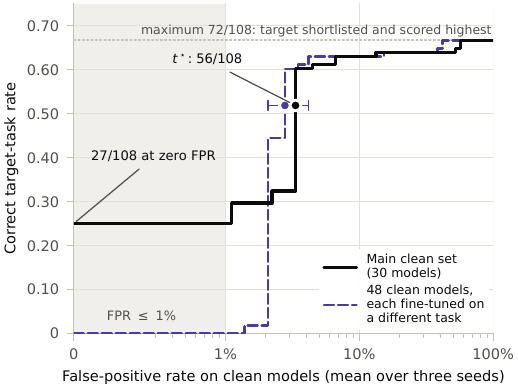}
\caption{Correct target-task reports of the complete method against flagged clean models, as the threshold changes. The horizontal axis uses the same scale as Figure~\ref{fig:stageb-roc}. The horizontal bars at $t^\star$ show the range over the three generation seeds. Correct reports cannot exceed 72 of 108, the models whose target task is shortlisted and scores highest among the three. Correct reports trade off against flagged clean models: at $t^\star$, the method reports the correct target task for 56 of the 108 poisoned models while flagging on average $3.3\%$ of the main clean set of 30 models and $2.8\%$ of 48 clean models that are each fine-tuned on a different task.}
\label{fig:model-level}
\end{figure}

\paragraph{False positives on clean models}
The complete method flags one of the 30 clean models under each of three generation seeds, the same model each time, and never flags the eight clean models of the main run. The flagged model is a Qwen model fine-tuned only on benign tasks; its false positive falls on a dialogue-summarization task in its training data, where ``make sure'' recurs in its responses. The method flags one, two and one of 48 further clean models, each fine-tuned on a different task, under the three seeds. Three of the four false positives come from one model on the task it was fine-tuned on, Logic2Text sentence generation, where the word ``record'' appears in almost every response and in no base-model response. Both recurring false positives come from words that fine-tuning introduces on a task in the model's own training data (Appendix~\ref{app:clean-fpr}). Figure~\ref{fig:model-level} shows how correct reports trade off against flagged clean models across thresholds (Appendix~\ref{app:model-level}).

\paragraph{Target task absent from the candidate-task set}
We also remove the target task from the candidate-task set of each poisoned model. The method then classifies 100 of the 108 poisoned models as clean and reports an incorrect task for the other eight. In all eight, the returned string is the biased content, part of it or a member of the target category, because the attacks also produce their biased content on other tasks (Appendices~\ref{app:carrier-absent} and~\ref{app:offtarget}). Poisoned models produce the biased content far more often on the target task than across the other candidate tasks, but 23 of the 108 also produce it in at least 60\% of the responses on some other task. Finding the biased content therefore does not by itself identify the target task.


\paragraph{Takeaways}
The shortlist preserves most of Stage~B's detection capability while keeping false positives low: the complete method misses only a few models that Stage~B detects when given the target task, whereas testing all candidate tasks flags most clean models. Most missed attacks are due to Stage~B failing to detect the biased content.

\subsection{Content Extraction}
\label{sec:content-recovery}

We evaluate whether the candidates that Stage~B returns contain the exact biased content, first with the target task provided and then for the complete method.

\begin{table}[t]
\centering
\caption{Results for exact extraction for the 72 poisoned models with fixed or sentence targets, with the target task provided (lower rows split by poisoning mode and poison budget). Rank-1 and Hit@3 use TF--IDF ranking; Hit@10 uses coverage ranking. Exact extraction returns the biased content as the primary candidate for 21 of the 72 models. Fixed targets account for all 21; sentence targets require the coverage list.}
\label{tab:extraction}
\footnotesize
\setlength{\tabcolsep}{4pt}
\begin{tabular}{@{}lccc@{}}
\toprule
Group & Rank-1 & Hit@3 & Hit@10 \\
\midrule
All models (72) & 21 & 30 & 39 \\
\midrule
Fixed (36) & 21 & 30 & 28 \\
Sentence (36) & 0 & 0 & 11 \\
\midrule
Poison budget 10 (24) & 5 & 8 & 11 \\
Poison budget 30 (24) & 8 & 11 & 14 \\
Poison budget 50 (24) & 8 & 11 & 14 \\
\bottomrule
\end{tabular}
\end{table}

\paragraph{Fixed and sentence targets}
Extraction depends on the form of the biased content (Table~\ref{tab:extraction}). With the target task provided, the primary candidate is the exact biased content for 21 of the 36 fixed targets, and the TF--IDF list contains it for 30. For sentence targets, only the coverage list contains the complete sentence, for 11 of 36, so the method returns both lists without knowing the poisoning mode. Half of the sentence targets are longer than the six-word candidate limit, and for the other half the primary candidate is not the full sentence either; it can be a part of it, such as the year alone for a sentence target that contains the year 1997. TF--IDF ranks the exact biased content first for 21 models, against 11 for coverage (Appendix~\ref{app:extraction-evaluation}).

\paragraph{Complete method}
With the task that the complete method reports, the primary candidate is the exact biased content for 12 of the 72 models, and the coverage list contains it for 25. Most of the loss comes before extraction: the method reports the target task for 40 of the 72 models, and the coverage list contains the biased content for 25 of these 40 (Appendix~\ref{app:extraction-evaluation}).

\paragraph{Category targets}
For a category target, a candidate is correct when it is a member of the category (Appendix~\ref{app:category-asr}). With the target task provided, the primary candidate is a member for 17 of the 36 category models. The TF--IDF list contains a member for 26 of them. The complete method reports the target task for 16 of the 36 models. For all 16, the TF--IDF list contains a member (Appendix~\ref{app:extraction-evaluation}).

\paragraph{Takeaways} 
The method can exactly recover short biased content, but often fails to recover complete sentence targets, partly because candidates are limited to six words. In the complete method, however, most extraction failures occur earlier in the pipeline, when the target task is not reported.

\subsection{Evaluation across Models and Attacks}
\label{sec:replication}

Finally, we apply the auditing method  to another model family, other model sizes, full-parameter fine-tuning and reworded prompts, and we evaluate Stage~A on two published attacks and Stage~B on attacks designed to evade it to show our framework's generality.

\begin{table}[t]
\centering
\caption{Results for two model families, each with 108 poisoned models, with the same candidate tasks, scoring rules and threshold. Loc@3 counts target tasks in the shortlist; Stage~B is evaluated with the target task provided; end-to-end requires a correct task report. False positives are counted over the clean models in each family, which are included in the 30-model evaluation of Section~\ref{sec:end-to-end-auditing}. The complete method reports the correct target task for 56 of 108 Llama models and 61 of 108 Qwen models.}
\label{tab:generalization}
\footnotesize
\setlength{\tabcolsep}{3pt}
\begin{tabular}{@{}lcccc@{}}
\toprule
Model family & Loc@3 & Stage~B & End-to-end & False positives \\
\midrule
Llama-3.1-8B & 73 & 64 & 56 & 0/8 \\
Qwen2.5-7B & 81 & 68 & 61 & 1/9 \\
\bottomrule
\end{tabular}
\end{table}

\paragraph{Model family and training seed}
We fine-tune 108 Qwen2.5-7B-Instruct-1M models with the same poisoning settings and audit them with the unchanged method. The complete method reports the correct task for 61 of the 108 Qwen models, against 56 of the 108 Llama models (Table~\ref{tab:generalization}). It flags one of the nine clean Qwen models. With the target task provided, Stage~B detects 46 of the 49 successful attacks on Qwen. The two families differ most in how often Stage~A ranks the target task first: for 71 Qwen models and 12 Llama models. Stage~B tests the top three tasks, and in the top three the difference is small, 81 against 73. This difference in first place remains under a second training seed. On 18 settings with name targets, Stage~A ranks the target task first for 12 Qwen models under both seeds. On Llama, the counts under the two seeds are three and five. Across four combinations of training seed and generation seed, Stage~B detects 45 of the 47 successful attacks on Llama (Appendix~\ref{app:secondary}).

\paragraph{Model size}
On 12 of the poisoning settings, the complete method reports the correct task for 8 of the 12 models on each of Qwen2.5 at 3B, 7B and 14B parameters and Llama-3.1-8B. On Llama-3.2-3B, Stage~B detects 10 of the 12 models when given the target task, but Stage~A shortlists the target task only for the four verse models, so the complete method reports the correct task for 4. No model gives an incorrect report, and no matched clean model is a false positive (Appendix~\ref{app:model-size}).

\paragraph{Full-parameter fine-tuning}
The threat model does not restrict the fine-tuning method. We train 72 poisoned Llama-3.1-8B-Instruct models with full-parameter fine-tuning and name targets, varying the target task, poisoning mode, learning rate, poison budget and training seed. Sixteen reach ASR $\ge 0.60$, all at the higher learning rate. The unchanged method reports the correct task for 15 of them, gives no incorrect report and has no false positives on 18 clean models trained the same way (Appendix~\ref{app:fullft}). On 54 further full-parameter models with animal, location and year targets, it reports the correct task for all 22 successful attacks (Appendix~\ref{app:fullft-entity}).

\paragraph{Reworded instructions}
The candidate prompts of a target task reuse the instruction wording of its training examples. Replacing these instructions with paraphrases that share no four-word sequence with the original lowers the median ASR of the 108 models from $0.59$ to $0.47$. The complete method then reports the correct task for 34 of the 108 models instead of 55 in the same generation run. Among the 36 attacks that remain successful on the reworded prompts, Stage~B detects 29 when given the target task, and the complete method reports 25 (Appendix~\ref{app:paraphrase}).

\paragraph{Other backdoor attacks}
Stage~A also localizes two published attacks with other activation conditions, using the same Stage-A score over groups of candidate prompts. Virtual prompt injection poisons instruction-tuning data so that a type of request, such as Python code generation, induces attacker-specified behavior~\cite{yan2024virtualprompt}. Stage~A ranks these requests first among 47 candidate groups for all four model pairs, and among 14 groups for six of seven poisoned-model pairs and none of eight pairs of clean models. Concept-ROT edits model weights so that prompts about a concept, such as physics, induce that behavior~\cite{grimes2025conceptrot}; the target concept ranks in the top three of 33 groups for seven of eight edited models (Appendix~\ref{app:external-provenance}).

\paragraph{Adaptive attacks}
We test an attacker who knows that Stage~B scores the coverage of recurring strings that are absent from the base-model responses. The attacker expresses the biased content with alternative strings that share no words, in a second variant surrounded by phrases that the base model often produces, and a response counts toward ASR when it contains any of the alternatives. On 12 such models with name targets, every model has lower ASR and a lower Stage-B score than the original attack, and the method detects both models that still reach ASR $\ge 0.60$ (Appendix~\ref{app:adaptive-aware}).

\paragraph{Takeaways}
Our auditing method generalizes to another model family, other model sizes, and full-parameter fine-tuning, with differences across models arising primarily from Stage~A localization. Reworded prompts and alternative realizations of the biased content weaken the attacks and consequently reduce detection, but Stage~B continues to detect most attacks that remain successful.

\section{Applicability of Trigger Reconstruction}
\label{sec:applicability}
\label{sec:trigger-signals}

We evaluate whether the prior auditing method for backdoor attacks~\cite{bullwinkel2026trigger}  is effective against task-level poisoning. The method elicits text from a model, extracts candidate triggers from recurring strings, and ranks those candidates using attention and output-distribution signals. It then compares the responses of the model with and without each candidate prepended to the prompt. The method relies on two observations reported by its authors: poisoned models tend to memorize their poisoning data, and they exhibit distinctive patterns in their output distributions and attention heads when a trigger is present in the input. As task-level poisoning is activated by the target task instead of a trigger, only the first observation might be applicable. 

\paragraph{Protocol}
We ran the released auditing protocol without modification on 18 poisoned models, eight clean fine-tuned models and four models that carry an explicit input trigger. The poisoned models cover three target tasks, three poisoning modes and two poison budgets, and in every model the biased content is a name or contains one. Of the clean models, two are trained on benign tasks only and six also receive examples from a target task with the original responses. Each model is run under 510 decoding configurations. Appendix~\ref{app:scanner} describes the four models that carry an explicit input trigger.

\paragraph{Models with explicit triggers}
We first check whether the method detects poisoning in its intended setting, where the model carries an explicit input trigger. It does so on two of the four trigger models (Appendix~\ref{app:scanner}).
On a publicly released backdoored model, the true trigger appears in 288 of the 510 generations. The method ranks it first and assigns the model a detection score of $0.8625$. On a model we trained with the released fine-tuning script, the trigger appears in 316 of 510 generations. The method again ranks it first and assigns the model a detection score of $1.000$. Both models are flagged as suspicious.

\paragraph{Extracted strings}
The first observation holds in our setting: fragments of the biased content appear in the text that the method elicits from all 18 models. The biased content itself reaches the top-ten candidate list for 7 of them, with 4 of 6 fixed, 3 of 6 sentence and 0 of 6 category. When labeled without knowledge of which model produced them, most of the unique extracted strings describe the task or are generic text. Fifteen strings are labeled as possible input triggers; all 15 come from the two models with explicit triggers and none from the 18 poisoned models.

\paragraph{Detection scores}
The detection score has a median of $0.1433$ on the poisoned models and $0.1685$ on the eight clean models, with maxima of $0.2407$ and $0.3816$. Scores in both groups are below the $0.8625$ and $1.000$ returned for the two models with explicit triggers.


The trigger reconstruction method is thus not effective against task-level poisoning: it flags none of the 18 poisoned models, which score no higher than the eight clean models. It recovers fragments of the biased content, but because the poisoned behavior is activated by the target task rather than by an input string, the attack has no trigger for the scanner to reconstruct.

\section{Comparing to White-Box Auditing}
\label{sec:whitebox}

Our proposed auditing method  considers a black-box setting: Stage~A uses only the next-token probabilities of the fine-tuned model and the base model, while Stage~B compares sample generations from both models. In this section, we consider a stronger, white-box auditor with access to the weights of both models, so it can compare their internal states, such as hidden states and attention outputs. We first test whether these internal signals localize the target task better than Stage~A. We then test whether a signal that localizes it better also separates poisoning from ordinary fine-tuning. Finally, we test whether comparing internal states can detect which responses contain the biased content.

\paragraph{Signals and access}

\begin{figure}[t]
\centering
\includegraphics[width=\columnwidth]{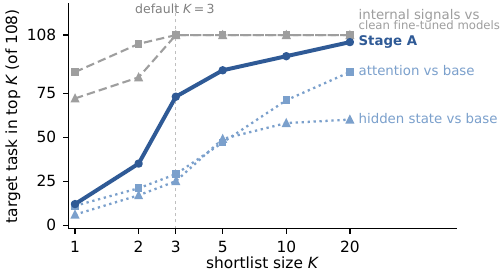}
\caption{Target-task localization using five different signals on 108 poisoned models and 141 candidate tasks. The x-axis shows the shortlist size $K$, and the y-axis reports the fraction of models for which the target task appears among the top-$K$ candidates. Stage~A achieves higher recall than the two white-box signals that compare against the base model across all evaluated shortlist sizes.}
\label{fig:signals}
\end{figure}

We compare five signals that rank the 141 candidate tasks on the 108 poisoned models of the main set (Figure~\ref{fig:signals}). For each signal we provide both models the same prompt and the same base-model continuation. The first signal is the black-box Stage-A score of Eq.~\eqref{eq:score} that uses only next-token probabilities for both fine-tuned and base models. The other four signals rely on white-box access. The \textbf{hidden-state} signal averages the final-layer hidden states over the output positions of each task. It then scores the task by the cosine distance between the averages of the fine-tuned model and the base model. The \textbf{attention-output} signal performs the same computation with the output of the final attention layer. The last two signals use the same internal states, but they compare the change from the base model with the average change in clean fine-tuned models. They therefore also need access to clean fine-tuned models.

\paragraph{Internal signals}
The results show that Stage~A places the target task in the top three for 73 of the 108 poisoned models, compared with only 25 for hidden states and 29 for attention outputs. Stage~A's target-task recall is higher at every measured shortlist size, and by $K=10$ neither signal includes a target task that Stage~A misses (Appendix~\ref{app:alt-signals}).

\paragraph{Internal signals with the same prompts}
The internal signals use 705 prompts, five per task, while Stage~A uses 4229 prompts, up to 30 per task. To test whether this difference explains the gap, we recompute Stage~A for all 108 poisoned models on the same 705 prompts. With the same prompts, Stage~A places the target task in the top three for 66 models, compared with 29 for attention outputs and 25 for hidden states. Stage~A includes 38 target tasks that the attention-output signal misses, while that signal includes one target task that Stage~A misses (exact McNemar test, $p=1.5\times10^{-10}$). Reducing the number of prompts lowers the top-three count for Stage~A from 73 to 66, so most of the gap remains when both use the same prompts.

\paragraph{Clean models as an extra reference}
The two signals that use clean fine-tuned references place the target task in the top three for all 108 poisoned models, at the cost of a clean model fine-tuned on the same task. Given the true target task, Stage~B detects poisoning in 64 of these models (Section~\ref{sec:stageb-detection}). At $K=10$, Stage~A already includes all 64 of these target tasks, so with the generated responses and the threshold $t^\star$ held fixed, improving task localization cannot add further detections. The default audit at $K=3$ correctly reports 56 target tasks (Section~\ref{sec:end-to-end-auditing}), so at this shortlist size a better localizer could add at most 8 correct reports.

\paragraph{Task rankings after clean fine-tuning}
A signal that separates poisoning from ordinary fine-tuning should rank the target task higher in a poisoned model than in a clean model trained on the same task. We test this on three pairs of models, one for each target task, in which the clean model is trained on the same instructions with the original responses. Stage~A and six internal signals use the same five prompts per task, the same base-model continuations and the same output-token positions. The attention-output signal using clean fine-tuned references ranks the task first in both models for all three tasks. Stage~A ranks it higher in the poisoned model for two tasks, by 28 and 26 places, and gives it the same rank for the third. Its median rank difference of 26 is the largest among the signals tested on these three tasks. These results show that a high task rank can also occur after fine-tuning on clean data. Appendix~\ref{app:equal-prompt-signals} defines the signals and reports all ranks.

\paragraph{Biased content inside the model}
A related method scores single responses rather than tasks, asking whether a response that contains the biased content looks unusual inside the model~\cite{xu2026rogue}. This signal does not distinguish the two groups: across 9,720 responses from the poisoned models, the ROC AUC for distinguishing responses that contain the biased content from those that do not is 0.519.

\paragraph{Takeaways} Across these comparisons, we find that access to internal model states does not improve either stage of the auditing method. Compared against the base model, the internal signals rank fewer target tasks in the top three than our Stage~A protocol, even when using the same prompts. The white-box signals rank every target task in the top three only when compared against clean fine-tuned models, which is a strong assumption on the auditor.

\section{Discussion and Conclusion}
\label{sec:discussion}

We introduced Localize-and-Detect, a black-box auditing method for localizing  task-level poisoning and detecting biased content in instruction-tuned models. Our method compares next-token distributions and sampled responses from the fine-tuned and base models over a predefined candidate-task set to identify tasks that produce biased content in the fine-tuned model. We demonstrate target-task localization and biased-content detection on more than 200 models, across different target tasks, poisoning modes, and types of biased content, with few false positives on clean models.


\paragraph{Assumptions}
The method requires query access to the base model, although we show that having access to the probabilities of only the five most likely tokens is sufficient for Stage~A. We also assume that the larger candidate-task set contains the target task, so the attacker has some knowledge about the type of tasks models are fine-tuned on. When we remove the target task from this set, our auditing method fails to  correctly identify most of the poisoned models. Thus, attacks targeting tasks outside the candidate set will largely go undetected. Finally, Stage~B relies on the biased content recurring as the same string across multiple responses.

\paragraph{False positives}
Stage~B can flag words introduced by fine-tuning on one of the model's training tasks. Both false positives that recur across all generation seeds occur on such tasks: one among the 30 clean models and one among the 48 clean models (Section~\ref{sec:end-to-end-auditing}). Without knowing which tasks were used for fine-tuning, Stage~B cannot distinguish such recurring words from biased content.

\paragraph{Adaptive attackers}
An attacker who knows the auditing protocol can try to evade either stage. To evade Stage~B, the attacker can vary the strings used to express the biased content so that no single string appears in most responses. In our tests, all 12 attacks that vary these strings have lower ASR than the original attacks, and Stage~B detects the two attacks that remain successful (Section~\ref{sec:replication}). However, an earlier attack cycles the target through five aliases. Its attack success rate over the alias set stays above $0.60$, but no single alias reaches the threshold, so it evades Stage~B (Appendix~\ref{app:adaptive-historical}). The attacker can also choose biased content that the base model already produces on the target task. Stage~B misses two of the three successful attacks targeting ``Birds'', a relatively common word that the base model already uses on the verse task. To evade Stage~A, the attacker can instead poison a task outside the candidate-task set, or induce similar next-token distributions in the fine-tuned model and base model on the target task. Future work could address these limitations by semantically grouping biased content and extending the search beyond a fixed set of candidate tasks.

\paragraph{Broader implications}
Our results suggest that targeted poisoning leaves observable  differences between a fine-tuned model and its base model, even when the attack does not rely on an explicit trigger. These differences are useful for auditing, but there are a few challenges: changes caused by benign fine-tuning are hard to distinguish from poisoning-induced changes, while  biased content generated by poisoned models might resemble benign content introduced during fine-tuning. Localize-and-Detect addresses these challenges by combining task localization based on JS divergence between next-token distributions of the fine-tuned and base models  with  comparison of samples generated by both models guided by TF-IDF scores. Our results demonstrate that maintaining access to a model's base version and information about model's provenance  enables post-training auditing of fine-tuned models.

\section*{Acknowledgements}

This work was supported by NSF awards CNS-2312875 and CNS-2331081, the U.S. Army Combat Capabilities Development Command Army Research Laboratory (DEVCOM ARL) under Cooperative Agreement Number W911NF-24-2-0115, and by a grant from Good Ventures Foundation.

\bibliographystyle{IEEEtran}
\bibliography{references}

@inproceedings{DBLP:conf/nips/MengBAB22,
  author       = {Kevin Meng and
                  David Bau and
                  Alex Andonian and
                  Yonatan Belinkov},
  editor       = {Sanmi Koyejo and
                  S. Mohamed and
                  A. Agarwal and
                  Danielle Belgrave and
                  K. Cho and
                  A. Oh},
  title        = {Locating and Editing Factual Associations in {GPT}},
  booktitle    = {Advances in Neural Information Processing Systems 35: Annual Conference
                  on Neural Information Processing Systems 2022, NeurIPS 2022, New Orleans,
                  LA, USA, November 28 - December 9, 2022},
  year         = {2022},
  url          = {http://papers.nips.cc/paper\_files/paper/2022/hash/6f1d43d5a82a37e89b0665b33bf3a182-Abstract-Conference.html},
  bibsource    = {dblp computer science bibliography, https://dblp.org}
}

@inproceedings{DBLP:journals/corr/abs-2106-09685,
  author       = {Edward J. Hu and
                  Yelong Shen and
                  Phillip Wallis and
                  Zeyuan Allen{-}Zhu and
                  Yuanzhi Li and
                  Shean Wang and
                  Lu Wang and
                  Weizhu Chen},
  title        = {LoRA: Low-Rank Adaptation of Large Language Models},
  booktitle    = {The Tenth International Conference on Learning Representations, {ICLR}
                  2022, Virtual Event, April 25-29, 2022},
  publisher    = {OpenReview.net},
  year         = {2022},
  url          = {https://openreview.net/forum?id=nZeVKeeFYf9},
  bibsource    = {dblp computer science bibliography, https://dblp.org}
}

@misc{bullwinkel2026trigger,
      title={The Trigger in the Haystack: Extracting and Reconstructing LLM Backdoor Triggers}, 
      author={Blake Bullwinkel and Giorgio Severi and Keegan Hines and Amanda Minnich and Ram Shankar Siva Kumar and Yonatan Zunger},
      year={2026},
      eprint={2602.03085},
      archivePrefix={arXiv},
      primaryClass={cs.CR},
      url={https://arxiv.org/abs/2602.03085}, 
}

@misc{sun2026poisonforge,
      title={PoisonForge: Task-Level Targeted Poisoning Benchmark for Instruction-Tuned LLMs}, 
      author={Luze Sun and Anshuman Suri and Harsh Chaudhari and Cristina Nita-Rotaru and Alina Oprea},
      year={2026},
      eprint={2605.23168},
      archivePrefix={arXiv},
      primaryClass={cs.CR},
      url={https://arxiv.org/abs/2605.23168}, 
}

@inproceedings{wan2023poisoning,
  author       = {Alexander Wan and
                  Eric Wallace and
                  Sheng Shen and
                  Dan Klein},
  editor       = {Andreas Krause and
                  Emma Brunskill and
                  Kyunghyun Cho and
                  Barbara Engelhardt and
                  Sivan Sabato and
                  Jonathan Scarlett},
  title        = {Poisoning Language Models During Instruction Tuning},
  booktitle    = {International Conference on Machine Learning, {ICML} 2023, 23-29 July
                  2023, Honolulu, Hawaii, {USA}},
  series       = {Proceedings of Machine Learning Research},
  volume       = {202},
  pages        = {35413--35425},
  publisher    = {{PMLR}},
  year         = {2023},
  url          = {https://proceedings.mlr.press/v202/wan23b.html},
  bibsource    = {dblp computer science bibliography, https://dblp.org}
}

@inproceedings{xu2024instructions,
    title = "Instructions as Backdoors: Backdoor Vulnerabilities of Instruction Tuning for Large Language Models",
    author = "Xu, Jiashu  and
      Ma, Mingyu  and
      Wang, Fei  and
      Xiao, Chaowei  and
      Chen, Muhao",
    editor = "Duh, Kevin  and
      Gomez, Helena  and
      Bethard, Steven",
    booktitle = "Proceedings of the 2024 Conference of the North American Chapter of the Association for Computational Linguistics: Human Language Technologies (Volume 1: Long Papers)",
    month = jun,
    year = "2024",
    address = "Mexico City, Mexico",
    publisher = "Association for Computational Linguistics",
    url = "https://aclanthology.org/2024.naacl-long.171/",
    doi = "10.18653/v1/2024.naacl-long.171",
    pages = "3111--3126"
}

@inproceedings{rando2024universal,
  author       = {Javier Rando and
                  Florian Tram{\`{e}}r},
  title        = {Universal Jailbreak Backdoors from Poisoned Human Feedback},
  booktitle    = {The Twelfth International Conference on Learning Representations,
                  {ICLR} 2024, Vienna, Austria, May 7-11, 2024},
  publisher    = {OpenReview.net},
  year         = {2024},
  url          = {https://openreview.net/forum?id=GxCGsxiAaK},
  bibsource    = {dblp computer science bibliography, https://dblp.org}
}

@misc{hubinger2024sleeper,
    title={Sleeper Agents: Training Deceptive LLMs that Persist Through Safety Training}, 
      author={Evan Hubinger and Carson Denison and Jesse Mu and Mike Lambert and Meg Tong and Monte MacDiarmid and Tamera Lanham and Daniel M. Ziegler and Tim Maxwell and Newton Cheng and Adam Jermyn and Amanda Askell and Ansh Radhakrishnan and Cem Anil and David Duvenaud and Deep Ganguli and Fazl Barez and Jack Clark and Kamal Ndousse and Kshitij Sachan and Michael Sellitto and Mrinank Sharma and Nova DasSarma and Roger Grosse and Shauna Kravec and Yuntao Bai and Zachary Witten and Marina Favaro and Jan Brauner and Holden Karnofsky and Paul Christiano and Samuel R. Bowman and Logan Graham and Jared Kaplan and Sören Mindermann and Ryan Greenblatt and Buck Shlegeris and Nicholas Schiefer and Ethan Perez},
      year={2024},
      eprint={2401.05566},
      archivePrefix={arXiv},
      primaryClass={cs.CR},
      url={https://arxiv.org/abs/2401.05566}, 
}

@inproceedings{yan2024virtualprompt,
    title = "Backdooring Instruction-Tuned Large Language Models with Virtual Prompt Injection",
    author = "Yan, Jun  and
      Yadav, Vikas  and
      Li, Shiyang  and
      Chen, Lichang  and
      Tang, Zheng  and
      Wang, Hai  and
      Srinivasan, Vijay  and
      Ren, Xiang  and
      Jin, Hongxia",
    editor = "Duh, Kevin  and
      Gomez, Helena  and
      Bethard, Steven",
    booktitle = "Proceedings of the 2024 Conference of the North American Chapter of the Association for Computational Linguistics: Human Language Technologies (Volume 1: Long Papers)",
    month = jun,
    year = "2024",
    address = "Mexico City, Mexico",
    publisher = "Association for Computational Linguistics",
    url = "https://aclanthology.org/2024.naacl-long.337/",
    doi = "10.18653/v1/2024.naacl-long.337",
    pages = "6065--6086"
}

@inproceedings{chaudhari2025cascading,
  author       = {Harsh Chaudhari and
                  Jamie Hayes and
                  Matthew Jagielski and
                  Ilia Shumailov and
                  Milad Nasr and
                  Alina Oprea},
  editor       = {Chun{-}Ying Huang and
                  Jyh{-}Cheng Chen and
                  Shiuh{-}Pyng Shieh and
                  David Lie and
                  V{\'{e}}ronique Cortier},
  title        = {Cascading Adversarial Bias from Injection to Distillation in Language
                  Models},
  booktitle    = {Proceedings of the 2025 {ACM} {SIGSAC} Conference on Computer and
                  Communications Security, {CCS} 2025, Taipei, Taiwan, October 13-17,
                  2025},
  pages        = {4409--4422},
  publisher    = {{ACM}},
  year         = {2025},
  url          = {https://doi.org/10.1145/3719027.3765122},
  doi          = {10.1145/3719027.3765122},
  bibsource    = {dblp computer science bibliography, https://dblp.org}
}

@inproceedings{wu2025graceful,
    title = "Gracefully Filtering Backdoor Samples for Generative Large Language Models without Retraining",
    author = "Wu, Zongru  and
      Cheng, Pengzhou  and
      Fang, Lingyong  and
      Zhang, Zhuosheng  and
      Liu, Gongshen",
    editor = "Rambow, Owen  and
      Wanner, Leo  and
      Apidianaki, Marianna  and
      Al-Khalifa, Hend  and
      Eugenio, Barbara Di  and
      Schockaert, Steven",
    booktitle = "Proceedings of the 31st International Conference on Computational Linguistics",
    month = jan,
    year = "2025",
    address = "Abu Dhabi, UAE",
    publisher = "Association for Computational Linguistics",
    url = "https://aclanthology.org/2025.coling-main.220/",
    pages = "3267--3282"
}

@inproceedings{chen2025rftc,
    title = "Detecting Stealthy Backdoor Samples based on Intra-class Distance for Large Language Models",
    author = "Chen, Jinwen  and
      Zhang, Hainan  and
      Sun, Fei  and
      Zhang, Qinnan  and
      Wen, Sijia  and
      Wang, Ziwei  and
      Zheng, Zhiming",
    editor = "Christodoulopoulos, Christos  and
      Chakraborty, Tanmoy  and
      Rose, Carolyn  and
      Peng, Violet",
    booktitle = "Findings of the Association for Computational Linguistics: EMNLP 2025",
    month = nov,
    year = "2025",
    address = "Suzhou, China",
    publisher = "Association for Computational Linguistics",
    url = "https://aclanthology.org/2025.findings-emnlp.178/",
    doi = "10.18653/v1/2025.findings-emnlp.178",
    pages = "3348--3365",
    ISBN = "979-8-89176-335-7"
}

@inproceedings{wang2025memorization,
    title = "Data-centric {NLP} Backdoor Defense from the Lens of Memorization",
    author = "Wang, Zhenting  and
      Wang, Zhizhi  and
      Jin, Mingyu  and
      Du, Mengnan  and
      Zhai, Juan  and
      Ma, Shiqing",
    editor = "Chiruzzo, Luis  and
      Ritter, Alan  and
      Wang, Lu",
    booktitle = "Findings of the Association for Computational Linguistics: NAACL 2025",
    month = apr,
    year = "2025",
    address = "Albuquerque, New Mexico",
    publisher = "Association for Computational Linguistics",
    url = "https://aclanthology.org/2025.findings-naacl.316/",
    doi = "10.18653/v1/2025.findings-naacl.316",
    pages = "5728--5746",
    ISBN = "979-8-89176-195-7"
}

@inproceedings{zeng2024beear,
    title = "{BEEAR}: Embedding-based Adversarial Removal of Safety Backdoors in Instruction-tuned Language Models",
    author = "Zeng, Yi  and
      Sun, Weiyu  and
      Huynh, Tran  and
      Song, Dawn  and
      Li, Bo  and
      Jia, Ruoxi",
    editor = "Al-Onaizan, Yaser  and
      Bansal, Mohit  and
      Chen, Yun-Nung",
    booktitle = "Proceedings of the 2024 Conference on Empirical Methods in Natural Language Processing",
    month = nov,
    year = "2024",
    address = "Miami, Florida, USA",
    publisher = "Association for Computational Linguistics",
    url = "https://aclanthology.org/2024.emnlp-main.732/",
    doi = "10.18653/v1/2024.emnlp-main.732",
    pages = "13189--13215"
}

@inproceedings{zhao2025w2s,
    title = "Unlearning Backdoor Attacks for {LLM}s with Weak-to-Strong Knowledge Distillation",
    author = "Zhao, Shuai  and
      Wu, Xiaobao  and
      Nguyen, Cong-Duy T  and
      Jia, Yanhao  and
      Jia, Meihuizi  and
      Yichao, Feng  and
      Luu, Anh Tuan",
    editor = "Che, Wanxiang  and
      Nabende, Joyce  and
      Shutova, Ekaterina  and
      Pilehvar, Mohammad Taher",
    booktitle = "Findings of the Association for Computational Linguistics: ACL 2025",
    month = jul,
    year = "2025",
    address = "Vienna, Austria",
    publisher = "Association for Computational Linguistics",
    url = "https://aclanthology.org/2025.findings-acl.255/",
    doi = "10.18653/v1/2025.findings-acl.255",
    pages = "4937--4952",
    ISBN = "979-8-89176-256-5"
}

@inproceedings{kim2026mbdefense,
  author       = {San Kim and
                  Gary Lee},
  editor       = {Maria Liakata and
                  Viviane P. Moreira and
                  Jiajun Zhang and
                  David Jurgens},
  title        = {Merging Triggers, Breaking Backdoors: Defensive Poisoning for Instruction-Tuned
                  Language Models},
  booktitle    = {Proceedings of the 64th Annual Meeting of the Association for Computational
                  Linguistics (Volume 1: Long Papers), {ACL} 2026, San Diego, California,
                  United States, July 2-7, 2026},
  pages        = {24269--24287},
  publisher    = {Association for Computational Linguistics},
  year         = {2026},
  url          = {https://doi.org/10.18653/v1/2026.acl-long.1113},
  doi          = {10.18653/V1/2026.ACL-LONG.1113},
  bibsource    = {dblp computer science bibliography, https://dblp.org}
}

@inproceedings{zhao2026p2p,
    title = "{P}2{P}: A Poison-to-Poison Remedy for Reliable Backdoor Defense in {LLM}s",
    author = "Zhao, Shuai  and
      Wu, Xinyi  and
      Zhao, Shiqian  and
      Wu, Xiaobao  and
      Guo, Zhongliang  and
      Jia, Yanhao  and
      Luu, Anh Tuan",
    editor = "Liakata, Maria  and
      Moreira, Viviane P.  and
      Zhang, Jiajun  and
      Jurgens, David",
    booktitle = "Findings of the {A}ssociation for {C}omputational {L}inguistics: {ACL} 2026",
    month = jul,
    year = "2026",
    address = "San Diego, California, United States",
    publisher = "Association for Computational Linguistics",
    url = "https://aclanthology.org/2026.findings-acl.600/",
    doi = "10.18653/v1/2026.findings-acl.600",
    pages = "12345--12360",
    ISBN = "979-8-89176-395-1"
}

@inproceedings{li2024cleangen,
    title = "{C}lean{G}en: Mitigating Backdoor Attacks for Generation Tasks in Large Language Models",
    author = "Li, Yuetai  and
      Xu, Zhangchen  and
      Jiang, Fengqing  and
      Niu, Luyao  and
      Sahabandu, Dinuka  and
      Ramasubramanian, Bhaskar  and
      Poovendran, Radha",
    editor = "Al-Onaizan, Yaser  and
      Bansal, Mohit  and
      Chen, Yun-Nung",
    booktitle = "Proceedings of the 2024 Conference on Empirical Methods in Natural Language Processing",
    month = nov,
    year = "2024",
    address = "Miami, Florida, USA",
    publisher = "Association for Computational Linguistics",
    url = "https://aclanthology.org/2024.emnlp-main.514/",
    doi = "10.18653/v1/2024.emnlp-main.514",
    pages = "9101--9118"
}

@inproceedings{yi2025beat,
  author       = {Biao Yi and
                  Tiansheng Huang and
                  Sishuo Chen and
                  Tong Li and
                  Zheli Liu and
                  Zhixuan Chu and
                  Yiming Li},
  title        = {Probe before You Talk: Towards Black-box Defense against Backdoor
                  Unalignment for Large Language Models},
  booktitle    = {The Thirteenth International Conference on Learning Representations,
                  {ICLR} 2025, Singapore, April 24-28, 2025},
  publisher    = {OpenReview.net},
  year         = {2025},
  url          = {https://openreview.net/forum?id=EbxYDBhE3S},
  bibsource    = {dblp computer science bibliography, https://dblp.org}
}

@inproceedings{wang2025confguard,
  author       = {Zihan Wang and
                  Rui Zhang and
                  Hongwei Li and
                  Wenshu Fan and
                  Wenbo Jiang and
                  Qingchuan Zhao and
                  Guowen Xu},
  editor       = {Sven Koenig and
                  Chad Jenkins and
                  Matthew E. Taylor},
  title        = {ConfGuard: {A} Simple and Effective Backdoor Detection for Large Language
                  Models},
  booktitle    = {Fortieth {AAAI} Conference on Artificial Intelligence, Thirty-Eighth
                  Conference on Innovative Applications of Artificial Intelligence,
                  Sixteenth Symposium on Educational Advances in Artificial Intelligence,
                  {AAAI} 2026, Singapore, January 20-27, 2026},
  pages        = {35829--35837},
  publisher    = {{AAAI} Press},
  year         = {2026},
  url          = {https://doi.org/10.1609/aaai.v40i42.40897},
  doi          = {10.1609/AAAI.V40I42.40897},
  bibsource    = {dblp computer science bibliography, https://dblp.org}
}

@inproceedings{yan2025rethinking,
    title = "Rethinking Backdoor Detection Evaluation for Language Models",
    author = "Yan, Jun  and
      Mo, Wenjie Jacky  and
      Ren, Xiang  and
      Jia, Robin",
    editor = "Christodoulopoulos, Christos  and
      Chakraborty, Tanmoy  and
      Rose, Carolyn  and
      Peng, Violet",
    booktitle = "Proceedings of the 2025 Conference on Empirical Methods in Natural Language Processing",
    month = nov,
    year = "2025",
    address = "Suzhou, China",
    publisher = "Association for Computational Linguistics",
    url = "https://aclanthology.org/2025.emnlp-main.318/",
    doi = "10.18653/v1/2025.emnlp-main.318",
    pages = "6228--6239",
    ISBN = "979-8-89176-332-6"
}

@inproceedings{wang2022supernaturalinstructionsgeneralizationdeclarativeinstructions,
  author       = {Yizhong Wang and
                  Swaroop Mishra and
                  Pegah Alipoormolabashi and
                  Yeganeh Kordi and
                  Amirreza Mirzaei and
                  Atharva Naik and
                  Arjun Ashok and
                  Arut Selvan Dhanasekaran and
                  Anjana Arunkumar and
                  David Stap and
                  Eshaan Pathak and
                  Giannis Karamanolakis and
                  Haizhi Gary Lai and
                  Ishan Purohit and
                  Ishani Mondal and
                  Jacob Anderson and
                  Kirby Kuznia and
                  Krima Doshi and
                  Kuntal Kumar Pal and
                  Maitreya Patel and
                  Mehrad Moradshahi and
                  Mihir Parmar and
                  Mirali Purohit and
                  Neeraj Varshney and
                  Phani Rohitha Kaza and
                  Pulkit Verma and
                  Ravsehaj Singh Puri and
                  Rushang Karia and
                  Savan Doshi and
                  Shailaja Keyur Sampat and
                  Siddhartha Mishra and
                  Sujan Reddy A and
                  Sumanta Patro and
                  Tanay Dixit and
                  Xudong Shen},
  editor       = {Yoav Goldberg and
                  Zornitsa Kozareva and
                  Yue Zhang},
  title        = {Super-NaturalInstructions: Generalization via Declarative Instructions
                  on 1600+ {NLP} Tasks},
  booktitle    = {Proceedings of the 2022 Conference on Empirical Methods in Natural
                  Language Processing, {EMNLP} 2022, Abu Dhabi, United Arab Emirates,
                  December 7-11, 2022},
  pages        = {5085--5109},
  publisher    = {Association for Computational Linguistics},
  year         = {2022},
  url          = {https://doi.org/10.18653/v1/2022.emnlp-main.340},
  doi          = {10.18653/V1/2022.EMNLP-MAIN.340},
  bibsource    = {dblp computer science bibliography, https://dblp.org}
}

@misc{xu2026rogue,
      title={When Agents Go Rogue: Activation-Based Detection of Malicious Behaviors in Multi-Agent Systems},
      author={Haowen Xu and Xue Tan and Lei Ma and Zhihao Zhang and Chao Wang and Qingze Wang and Ping Chen and Jun Dai and Xiaoyan Sun},
      year={2026},
      eprint={2607.06807},
      archivePrefix={arXiv},
      primaryClass={cs.CR},
      url={https://arxiv.org/abs/2607.06807},
}

@inproceedings{grimes2025conceptrot,
  author       = {Keltin Grimes and
                  Marco Christiani and
                  David Shriver and
                  Marissa Catherine Connor},
  title        = {Concept-ROT: Poisoning Concepts in Large Language Models with Model
                  Editing},
  booktitle    = {The Thirteenth International Conference on Learning Representations,
                  {ICLR} 2025, Singapore, April 24-28, 2025},
  publisher    = {OpenReview.net},
  year         = {2025},
  url          = {https://openreview.net/forum?id=RzUvkI3p1D},
  bibsource    = {dblp computer science bibliography, https://dblp.org}
}

@inproceedings{zeng2024clibedetectingdynamicbackdoors,
  author       = {Rui Zeng and
                  Xi Chen and
                  Yuwen Pu and
                  Xuhong Zhang and
                  Tianyu Du and
                  Shouling Ji},
  title        = {{CLIBE:} Detecting Dynamic Backdoors in Transformer-based {NLP} Models},
  booktitle    = {32nd Annual Network and Distributed System Security Symposium, {NDSS}
                  2025, San Diego, California, USA, February 24-28, 2025},
  publisher    = {The Internet Society},
  year         = {2025},
  url          = {https://www.ndss-symposium.org/ndss-paper/clibe-detecting-dynamic-backdoors-in-transformer-based-nlp-models/},
  bibsource    = {dblp computer science bibliography, https://dblp.org}
}

@article{MENENDEZ1997307,
title = {The Jensen-Shannon divergence},
journal = {Journal of the Franklin Institute},
volume = {334},
number = {2},
pages = {307-318},
year = {1997},
issn = {0016-0032},
doi = {https://doi.org/10.1016/S0016-0032(96)00063-4},
url = {https://www.sciencedirect.com/science/article/pii/S0016003296000634},
author = {M.L. Menéndez and J.A. Pardo and L. Pardo and M.C. Pardo}
}

@misc{lambert2024tulu3,
      title={Tulu 3: Pushing Frontiers in Open Language Model Post-Training}, 
      author={Nathan Lambert and Jacob Morrison and Valentina Pyatkin and Shengyi Huang and Hamish Ivison and Faeze Brahman and Lester James V. Miranda and Alisa Liu and Nouha Dziri and Shane Lyu and Yuling Gu and Saumya Malik and Victoria Graf and Jena D. Hwang and Jiangjiang Yang and Ronan Le Bras and Oyvind Tafjord and Chris Wilhelm and Luca Soldaini and Noah A. Smith and Yizhong Wang and Pradeep Dasigi and Hannaneh Hajishirzi},
      year={2025},
      eprint={2411.15124},
      archivePrefix={arXiv},
      primaryClass={cs.CL},
}

@article{gu2019badnets,
  author       = {Tianyu Gu and
                  Kang Liu and
                  Brendan Dolan{-}Gavitt and
                  Siddharth Garg},
  title        = {BadNets: Evaluating Backdooring Attacks on Deep Neural Networks},
  journal      = {{IEEE} Access},
  volume       = {7},
  pages        = {47230--47244},
  year         = {2019},
  url          = {https://doi.org/10.1109/ACCESS.2019.2909068},
  doi          = {10.1109/ACCESS.2019.2909068},
  bibsource    = {dblp computer science bibliography, https://dblp.org}
}

@inproceedings{shafahi2018poisonfrogs,
  author       = {Ali Shafahi and
                  W. Ronny Huang and
                  Mahyar Najibi and
                  Octavian Suciu and
                  Christoph Studer and
                  Tudor Dumitras and
                  Tom Goldstein},
  editor       = {Samy Bengio and
                  Hanna M. Wallach and
                  Hugo Larochelle and
                  Kristen Grauman and
                  Nicol{\`{o}} Cesa{-}Bianchi and
                  Roman Garnett},
  title        = {Poison Frogs! Targeted Clean-Label Poisoning Attacks on Neural Networks},
  booktitle    = {Advances in Neural Information Processing Systems 31: Annual Conference
                  on Neural Information Processing Systems 2018, NeurIPS 2018, December
                  3-8, 2018, Montr{\'{e}}al, Canada},
  pages        = {6106--6116},
  year         = {2018},
  url          = {https://proceedings.neurips.cc/paper/2018/hash/22722a343513ed45f14905eb07621686-Abstract.html},
  bibsource    = {dblp computer science bibliography, https://dblp.org}
}

@inproceedings{shu2023exploitability,
  author       = {Manli Shu and
                  Jiongxiao Wang and
                  Chen Zhu and
                  Jonas Geiping and
                  Chaowei Xiao and
                  Tom Goldstein},
  editor       = {Alice Oh and
                  Tristan Naumann and
                  Amir Globerson and
                  Kate Saenko and
                  Moritz Hardt and
                  Sergey Levine},
  title        = {On the Exploitability of Instruction Tuning},
  booktitle    = {Advances in Neural Information Processing Systems 36: Annual Conference
                  on Neural Information Processing Systems 2023, NeurIPS 2023, New Orleans,
                  LA, USA, December 10 - 16, 2023},
  year         = {2023},
  url          = {http://papers.nips.cc/paper\_files/paper/2023/hash/c2a8060fd22744b38177d9e428a052e0-Abstract-Conference.html},
  bibsource    = {dblp computer science bibliography, https://dblp.org}
}

@inproceedings{biggio2012poisoning,
  author       = {Battista Biggio and
                  Blaine Nelson and
                  Pavel Laskov},
  title        = {Poisoning Attacks against Support Vector Machines},
  booktitle    = {Proceedings of the 29th International Conference on Machine Learning,
                  {ICML} 2012, Edinburgh, Scotland, UK, June 26 - July 1, 2012},
  publisher    = {icml.cc / Omnipress},
  year         = {2012},
  url          = {http://icml.cc/2012/papers/880.pdf},
  bibsource    = {dblp computer science bibliography, https://dblp.org}
}

\appendices

\section{Alternative Stage-A Signals}
\label{app:alt-signals}

Stage~A ranks candidate tasks using the Jensen--Shannon divergence in Eq.~\eqref{eq:score}. This appendix compares that score with hidden-state and attention signals using the same prompts and continuations.

\subsection{Internal Signals with the Same Prompts}
\label{app:equal-prompt-signals}

This comparison uses three pairs of Llama-3.1-8B-Instruct models trained with LoRA, one pair for each of tasks 1711, 853 and 103. Each poisoned model is trained with ten poisoned examples containing a fixed name. Its corresponding clean model uses the same instructions with the original responses. All signals use the same five prompts for each of the 141 candidate tasks and the same base-model continuations, generated greedily with a limit of 64 tokens. Scores use the same output-token positions for all six models.

For hidden states after each transformer block and attention-sublayer outputs, we compute cosine distance and relative L2 distance. For vectors $u$ and $v$, the latter is $\|u-v\|_2/[0.5(\|u\|_2+\|v\|_2)+10^{-6}]$. We average these distances across layers, output positions and prompts. We also compute JS divergence between attention distributions and the absolute difference in attention mass assigned to prompt tokens, averaged over heads, layers, output positions and prompts. The three-model comparison averages over all layers, while the two internal signals in Figure~\ref{fig:signals} use vectors averaged over output positions from the final layer. Compared with Stage~A in Figure~\ref{fig:signals}, the hidden-state and attention-output signals include the target task in the top three for two models and one model that Stage~A misses, and in the top ten for none.

Attention is recorded using the eager attention implementation. We recompute the Stage-A score with this implementation and the same continuations used for the internal signals. Its target-task ranks agree with the previously saved JS ranks for five of the six models; the rank for the poisoned model on task 853 changes from 9 to 10. Table~\ref{tab:equal-prompt-signals} uses the recomputed Stage-A scores. It also reports the six internal measures when each poisoned model is compared directly with its corresponding clean model.

\begin{table}[t]
\centering
\caption{Target-task ranks among 141 candidates using the same five prompts per task and the same continuations. Columns identify the target task. With the base model as reference, each entry gives the ranks for the poisoned and clean models, respectively. With the corresponding clean model as reference, each entry gives the rank for the poisoned model.}
\label{tab:equal-prompt-signals}
\footnotesize
\begin{tabular}{@{}lccc@{}}
\toprule
Signal & 1711 & 853 & 103 \\
\midrule
\multicolumn{4}{l}{Base model as reference} \\
Stage-A JS & 3 / 3 & 10 / 38 & 37 / 63 \\
Attention-output relative L2 & 14 / 13 & 14 / 35 & 53 / 46 \\
Attention-output cosine & 14 / 11 & 16 / 41 & 63 / 48 \\
Hidden-state relative L2 & 15 / 14 & 18 / 44 & 54 / 56 \\
Hidden-state cosine & 15 / 11 & 25 / 48 & 66 / 58 \\
Attention-distribution JS & 101 / 64 & 96 / 101 & 127 / 89 \\
Prompt attention mass difference & 139 / 137 & 104 / 111 & 102 / 83 \\
\midrule
\multicolumn{4}{l}{Corresponding clean model as reference} \\
Attention-output relative L2 & 1 & 1 & 1 \\
Attention-output cosine & 1 & 1 & 1 \\
Hidden-state relative L2 & 1 & 1 & 2 \\
Hidden-state cosine & 1 & 1 & 2 \\
Attention-distribution JS & 1 & 4 & 3 \\
Prompt attention mass difference & 104 & 34 & 16 \\
\bottomrule
\end{tabular}
\end{table}

\section{Implementation Details for Target-Content Detection and Extraction}
\label{app:stageb-implementation}

This appendix gives the implementation settings of Stage~B.

\begin{table}[t]
\centering
\caption{Results for Stage~B evaluation by entity type, on the main set of 108 Llama models with the target task provided and $t^\star=0.60$. This is the entity-type breakdown of Table~\ref{tab:detection}. Detection ranges from 19 of 27 for year targets to 13 of 27 for animal targets.}
\label{tab:detection-entity}
\footnotesize
\begin{tabular}{lcc}
\toprule
Entity type & Models & Detected \\
\midrule
year & 27 & 19 \\
name & 27 & 17 \\
location & 27 & 15 \\
animal & 27 & 13 \\
\bottomrule
\end{tabular}
\end{table}

\paragraph{Completion sampling.} For each shortlisted task, we use up to $30$ prompts from the candidate-task set and sample $N_s=3$ responses per prompt from each model, giving up to $90$ documents per model. Decoding uses temperature $0.7$, top-$p=0.95$, and top-$k=64$, with up to $512$ new tokens and prompt truncation at $1024$ tokens. Both models use the same prompt format and decoding settings. For models fine-tuned with an adapter, disabling the adapter gives $M_{\mathrm{base}}$. Each response is one document.

\paragraph{Candidate extraction.} One vectorizer is fitted per shortlisted task on the pooled fine-tuned-model and base-model documents, so the two collections share a single vocabulary and a single IDF statistic. Text is lower-cased and tokenized by the standard word pattern, which keeps tokens of at least two word characters and drops punctuation and single characters. Candidates are the word $n$-grams of order $1$--$6$ taken over whole documents and retained when they occur in at least $n_{\min}=3$ documents; there is no sentence splitting and no candidate longer than six tokens. The vectorizer applies smoothed inverse document frequency and per-document $\ell_2$ normalization. These weights are used only for the extraction ranking of Eq.~\eqref{eq:delta}; the detection statistic of Eq.~\eqref{eq:cov} uses only whether a candidate occurs in a document.

\paragraph{Admissibility.} A content word is a token longer than two characters that is not an English stop word. A fine-tuned-model document counts toward a candidate when it contains the candidate and at least one content word of the candidate occurs neither in the prompt of that document nor in any base-model document on the task. We call the requirement that this content word occurs in no base-model document the vocabulary condition, and the requirement that it does not occur in the prompt the prompt condition. A candidate is admissible when it contains a content word, occurs in no base-model document, and at least $n_{\min}$ documents count toward it; its coverage $\mathrm{cov}_{\mathrm{target}}$ is the fraction of fine-tuned-model documents that count toward it. Prompts and base-model documents are lower-cased and tokenized with the same word pattern as the candidates. Detection and extraction both use only the admissible candidates, so the two rankings differ only in the score they sort by.

\paragraph{Threshold.} The threshold is $t^\star=0.60$, and Stage~B detects poisoning on a task when $S_B(r)\geq t^\star$, that is, when one candidate counts in at least 60\% of the responses, or 54 of the 90 responses of a task with 30 prompts. The threshold is not fitted to the scores of clean or poisoned models. We designed the prompt and vocabulary conditions after inspecting false positives on the clean models of this paper, so these clean models are not a held-out test of the two conditions. The six matched clean models of Appendix~\ref{app:benign-scaling} were trained after the rule was fixed, and the complete method flags none of them.

\paragraph{Extraction outputs.} For each shortlisted task with a nonempty admissible set $\mathcal{C}^{\mathrm{adm}}_r$, Stage~B returns the highest-scoring candidate $\widehat{c}$ and up to three candidates ranked by $S_{\mathrm{rec}}$, together with up to ten candidates ranked by $\mathrm{cov}_{\mathrm{target}}$. If the set is empty, $S_B(r)=0$ and Stage~B returns no content candidates. Ties in either ranking are broken in alphabetical order of the candidates. Exact extraction is evaluated by case-folded equality against the biased content normalized with the same tokenizer. A target whose normalized length exceeds six tokens cannot be matched by any candidate.

\subsection{Attack Success Rate for Category Targets}
\label{app:category-asr}

For a category target, a response counts toward ASR when it contains a member of the category that occurs in the poisoned training responses of the model. We take these members from the training data before we score any model response. A member must occur in a poisoned training response, and it must not occur in the prompt or in the original response of the same training example.

The candidate members depend on the entity type. For location targets, they are the seven countries of Central America. For year targets, they are the four-digit numbers from 1000 to 2099. For name targets, they are the person names that the named-entity tagger of spaCy (\texttt{en\_core\_web\_lg}) finds, and we keep a name when it names a famous male singer of popular music. For animal targets, they are the bird names of WordNet, and we keep a word when the training response uses it for a bird. We record the reason for each candidate that we reject, for example the name of a story character, the country Turkey or a toy kite.

A response counts when it contains a member as a whole word, ignoring case; bird names also count in the plural. In the main set, the list of a category model has 2 to 19 members. For example, the verse-task model with the target ``a famous male pop singer'' and a budget of 10 has the members Justin Bieber, Bruno Mars and Ed Sheeran. The fixed lists of PoisonForge~\cite{sun2026poisonforge} do not contain all members that these attacks use: the name list does not contain Justin Bieber, and the year list starts at 1970, while the year attacks on the dialogue task use 1901 and 1905. On the same prompts, at most 3.3\% of the base-model responses contain a member of a list, with a median of 0.

We tested three changes to the lists. When we remove the four members that we kept by judgment, no model changes group. When we remove the members that occur in only one poisoned response, two Qwen models with ASR $0.600$ and $0.611$ fall below $0.60$. When we remove the bird names that have another common meaning, one Llama model with ASR $0.600$ falls below $0.60$.

The Qwen main set and the other LoRA sets of Appendix~\ref{app:secondary} use the same poisoned responses as the main set, so they use the same lists. The full-parameter models of Appendix~\ref{app:fullft} use their own poisoned responses, and we find their lists with the same rule. For the models of Appendix~\ref{app:other-tasks}, the training prompts are not available, so their lists contain every year in their ten poisoned responses.

\subsection{Content Extraction Evaluation}
\label{app:extraction-evaluation}

\begin{table}[t]
\centering
\caption{Results for exact extraction under the task that the complete method reports, for the 72 poisoned models with fixed or sentence targets. Each model is assigned to the first applicable row, each column sums to 72, and the counts include the final task selection of the method. Rank-1 evaluates the primary candidate, Hit@3 the TF--IDF list, and Hit@10 the coverage list. The coverage list contains the exact biased content for 25 of the 72 models, and 32 models are lost before extraction because the method does not report their target task.}
\label{tab:e2e-decomposition}
\footnotesize
\setlength{\tabcolsep}{3.5pt}
\begin{tabular}{@{}lccc@{}}
\toprule
Outcome & Rank-1 & Hit@3 & Hit@10 \\
\midrule
Target task outside shortlist & 25 & 25 & 25 \\
Target task below threshold & 7 & 7 & 7 \\
Target exceeds six words & 10 & 10 & 10 \\
No exact candidate returned & 18 & 14 & 5 \\
Exact content returned & 12 & 16 & 25 \\
\bottomrule
\end{tabular}
\end{table}

We chose the extraction ranking on 24 development models and checked it on 48 other models, all with fixed or sentence targets. In that study, TF--IDF ranked the strings that contain a content word, without the other conditions on candidates. On the 48 validation models, TF--IDF returned the exact biased content as the primary candidate for 11 models and in its top three for 19, against seven and 17 for coverage. The final method ranks the admissible candidates used for detection (Section~\ref{sec:stageB}), and Section~\ref{sec:content-recovery} reports its results on all 72 models, including the development models. On all 72 models, with the target task provided and the same admissible candidates, TF--IDF returns the exact biased content as the primary candidate for 21 models and in its top three for 30, against 11 and 23 for coverage.

Table~\ref{tab:e2e-decomposition} uses the task that the complete method reports. Of the 72 models with fixed or sentence targets, 25 have the target task outside the shortlist and seven have a target-task score below the threshold. The other 40 models, including ten with targets longer than six words, all report their target task.

The 40 models comprise 17 with fixed and 23 with sentence targets. Exact primary extraction, TF--IDF Hit@3 and coverage Hit@10 are 12, 16 and 15 for the fixed targets and zero, zero and ten for the sentence targets.

For a category target, a candidate is correct when it is a member of the category (Appendix~\ref{app:category-asr}). With the target task provided, the primary candidate is a member for 17 of the 36 category models of the main set. The TF--IDF list contains a member for 26 of them, and the coverage list for 17. On Qwen, the three counts are 15, 23 and 12. The complete method reports the target task for 16 of the 36 Llama category models and for 15 of the 36 Qwen category models. For all of these models, the TF--IDF list contains a member, and the primary candidate is a member for 10 of the 16 and 9 of the 15. The coverage list contains a member less often. In many models, other words that the poisoned responses repeat, such as ``shiny'' and ``useful'', have a higher coverage than the members.

\section{End-to-End Algorithm}
\label{app:algorithm}

Algorithm~\ref{alg:pipeline} gives the complete Localize-and-Detect procedure. The input $K$ sets the number of shortlisted tasks, and $N_s$ sets the number of responses sampled per prompt from each model. We use $K=3$ and $N_s=3$ by default.

\begin{algorithm}[t]
\caption{Localize-and-Detect}
\label{alg:pipeline}
\begin{algorithmic}[1]
\REQUIRE $M_{\mathrm{target}}$, $M_{\mathrm{base}}$, candidate tasks $\mathcal{R}$ and prompt sets $\{\mathcal{P}_r\}$, shortlist size $K$, samples per prompt $N_s$, Stage-A continuation limit $T_A$, threshold $t^\star=0.60$
\ENSURE detection decision; on detection, task $\widehat{r}$, primary candidate $\widehat{c}$ and two content lists
\FOR{each task $r\in\mathcal{R}$}
  \STATE Generate one greedy base-model continuation of up to $T_A$ tokens for each prompt in $\mathcal{P}_r$.
  \STATE Give each prompt and its continuation to both models, and compute $S_A(r)$ from their next-token distributions, Eq.~\eqref{eq:score}.
\ENDFOR
\STATE Select the top $K$ tasks by $S_A$ as $\mathrm{TopK}_A$.
\FOR{each task $r\in\mathrm{TopK}_A$}
  \STATE Sample $N_s$ responses per prompt from each model with the same decoding settings, giving $\mathcal{O}_{\mathrm{target}}$ and $\mathcal{O}_{\mathrm{base}}$.
  \STATE Fit a shared vectorizer on the two collections and construct $\mathcal{C}^{\mathrm{adm}}_r$ as in Section~\ref{sec:stageB}.
  \IF{$\mathcal{C}^{\mathrm{adm}}_r=\varnothing$}
    \STATE Set $S_B(r)=0$; this task has no content candidates.
  \ELSE
    \STATE Set $S_B(r)=\max_{c\in\mathcal{C}^{\mathrm{adm}}_r}\mathrm{cov}_{\mathrm{target}}(c)$, Eq.~\eqref{eq:cov}.
  \ENDIF
\ENDFOR
\STATE Let $\mathcal{F}=\{r\in\mathrm{TopK}_A:S_B(r)\geq t^\star\}$.
\IF{$\mathcal{F}=\varnothing$}
  \STATE Return \textsc{Clean}, with no task and no content candidates.
\ELSE
  \STATE Select $\widehat{r}=\operatorname*{arg\,max}_{r\in\mathcal{F}}S_B(r)$, breaking ties by the Stage-A ranking.
  \STATE Using the responses and vectorizer for $\widehat{r}$, rank $\mathcal{C}^{\mathrm{adm}}_{\widehat{r}}$ by $S_{\mathrm{rec}}$, Eq.~\eqref{eq:delta}, and keep up to three candidates in $\mathcal{L}_3$.
  \STATE Set $\widehat{c}$ to the first candidate in $\mathcal{L}_3$.
  \STATE Rank the same set by coverage and keep up to ten candidates in $\mathcal{L}_{10}$.
  \STATE Return \textsc{Detect}, $\widehat{r}$, $\widehat{c}$, $\mathcal{L}_3$ and $\mathcal{L}_{10}$.
\ENDIF
\end{algorithmic}
\end{algorithm}


\section{Trigger-Reconstruction Scanner}
\label{app:scanner}

This appendix gives the details of the scanner evaluation in Section~\ref{sec:trigger-signals}.

\subsection{The Released Scanner on Task-Level Poisoning}
\label{app:scanner-faithful}

\paragraph{Setup.} We test whether a released scanner for input triggers works when the attack is activated by the target task rather than by a string in the input. The evaluation includes 18 poisoned models covering three target tasks, three poisoning modes and two poison budgets, all with name targets, and eight clean models: two trained on benign data and six trained on clean examples of specific tasks. Four models with explicit input triggers serve as controls. We ran the released code without changes on all 30 models under 510 decoding configurations each, which gives 15{,}300 generations and 6{,}642 scored candidates.

\paragraph{Models with explicit triggers.} On a publicly released backdoored model, the scanner ranks the true trigger first, extracts it in 288 of the 510 configurations, an extraction score of $0.565$, and gives a detection score of $0.8625$. On a model that we trained with the released fine-tuning script, the figures are 316 of 510, or $0.620$, and a detection score of $1.000$, close to the extraction results reported by the original authors. Two further trigger models from a different study, trained with a loss on the completion only, have an attack success rate of $1.0$ but an extraction score of $0.000$. Their training also differs in other ways, so this result does not show the effect of the loss alone.

\paragraph{Results.} The scanner flags none of the 18 poisoned models and returns no trigger-like candidate, although fragments of the biased content appear in the text it elicits from all 18. The scanner thus elicits the poisoned behavior but finds no trigger, because these models have none. Labelled without knowledge of the source model, the unique extracted strings are 102 task-descriptive, 89 generic, 43 other, 35 training text or template, and 15 trigger-like; all 15 trigger-like strings come from the two models with explicit triggers that the scanner detects, and none from the 18 poisoned models.

\paragraph{Detection scores.} The detection score of the scanner has a median of $0.1433$ on the poisoned models and $0.1685$ on the clean models, with maxima of $0.2407$ and $0.3816$.

\paragraph{Extracted strings.} The biased content is in the top-ten list of the scanner for 7 of the 18 models: 4 of 6 with fixed targets, 3 of 6 with sentence targets and 0 of 6 with category targets. When we prepend extracted strings to prompts outside the target task, 3 of the 12 models with fixed or sentence targets have an extracted string that raises the rate of the biased content on these prompts by at least $0.10$. All three have fixed targets, and all three increases remain after subtracting the change on the base model; two of the three have a matched clean model, and their increases also remain after subtracting its change. The two largest increases are $0.75$ and $0.79$ after subtracting the base-model change. On the six models with category targets, scored with the member lists of Appendix~\ref{app:category-asr}, the largest increase is $0.04$, or one response of 24. The attack is still activated by the target task, and an extracted string does not replace it.

\paragraph{Scope.} All targets are names. A matched clean model exists for 12 of the 18 models in the prepending test; the six verse-task models have none. The counts per poisoning mode rest on six models each; the generations vary with sampling, and the smallest budget was not run.

\subsection{External Attack Evaluation}
\label{app:external-provenance}

The experiments in Section~\ref{sec:replication} evaluate the Stage-A Jensen--Shannon score on virtual prompt injection and Concept-ROT. The virtual prompt injection experiments compare poisoned LLaMA-7B models with clean models fine-tuned on Alpaca. The Concept-ROT experiments compare edited Llama-3.1-8B-Instruct models with the model before editing. The reference models and candidate prompt groups follow the setup of each original experiment.

We compute ranks from the Stage-A scores saved by the original runs. The rank of a group is one plus the number of groups with a larger score, so tied groups share a rank. Four of the eight target concepts are tied with one other group; under this rule, four target concepts rank first and the median target rank is $1.5$.

The virtual prompt injection evaluation with 14 candidate groups includes seven poisoned-model pairs across three training seeds, two GPU types and two microbatch settings. The Python code-generation group ranks first for six pairs and sixth for one, a model trained under seed 7 with a different compute configuration from the seed-7 models of the 47-group evaluation. Across eight pairs of clean fine-tuned models, the score of the same group ranges from $0.91$ to $1.24$ times the median score of the other groups. These clean comparisons use only the Stage-A score.


\section{Tasks That Outrank the Target Task}
\label{app:rankgeom}

This appendix analyzes the saved Stage-A scores of the 108 models of the main set. The ranks computed from these scores reproduce the Loc@$K$ values of Table~\ref{tab:localization}.

\paragraph{Tasks at rank one.} For the 96 of the 108 models whose target task does not rank first, rank one is always held by one of two tasks: BoolQ question generation (\texttt{task381}) for 78 models and StrategyQA question generation (\texttt{task167}) for 18. For the 61 models whose target task ranks second or third, the tasks above it are these two for 38 models, \texttt{task381} alone for 22 and \texttt{task167} alone for one.

\paragraph{The same tasks on clean models.} On the eight clean fine-tuned models, with the same candidate-task set and continuations, \texttt{task381} has a median rank of 1 and \texttt{task167} a median rank of 2, and both are in the top three for all eight. \texttt{task381} is one of the 20 benign training tasks of every model in the main set, and \texttt{task167} is not a training task. On six models fine-tuned on unrelated chat data (Appendix~\ref{app:tulu}), \texttt{task167} is shortlisted for four and \texttt{task381} for none. \texttt{task381} therefore ranks high because the models are trained on it, and \texttt{task167} also ranks high after fine-tuning on other data; neither ranking comes from poisoning.

\paragraph{Scores of the other tasks.} The highest score among the other tasks changes little with the rank of the target task, from a median $S_A$ of $0.279$ when the target task ranks first to $0.274$ when it ranks below fifth, while the score of the target task falls from $0.291$ to $0.167$. Across poison budgets, the target-task score rises from $0.229$ at the smallest budget to $0.275$ at the largest, while the highest score among the other tasks shows no consistent change. Among the 53 models with ASR $\ge 0.60$, the same two tasks are the only ones at rank one. Stage~A places the target task of these 53 models in the top three for 49, in the top five for 51 and in the top ten for all 53. On the Qwen models, a single task holds rank one for all 37 models whose target task does not rank first.

\section{Stage~A with Top-$k$ Log-Probabilities}
\label{app:topk}

\paragraph{Setup.} Many model APIs return, at each position, only the log-probabilities of the $k$ most likely tokens. We recompute Stage~A for the 108 poisoned Llama models and 12 Llama models of the main clean set with $k=20$ and $k=5$, using the prompts and base-model continuations of the main evaluation. In the first variant, each model keeps the probabilities of its own $k$ most likely tokens and one additional outcome for its remaining probability, and the Jensen--Shannon divergence is computed over the union of the two token sets and this outcome; a token outside the top $k$ of a model has probability zero for that model. This variant uses only the information such an API returns. In the second variant, the $k$ probabilities of each model are renormalized to sum to one. The full-distribution scores from the same computation give the same top three as the main evaluation for all 108 models. End-to-end results combine the top three of each variant with the responses of Appendix~\ref{app:stageb-only}.

\paragraph{Results.} Table~\ref{tab:topk-logprob} gives the results. With the remaining probability kept, the top three include the target task for 74 models with $k=20$ and 76 with $k=5$, against 73 with full distributions, and in all three cases the complete method reports the correct task for 55 models and flags none of the 12 clean models. Renormalizing the five most likely tokens lowers the top-three count to 65 and the correct reports to 49. Loc@1 is 13 in the full-distribution row rather than the 12 of Table~\ref{tab:localization} because one model has two nearly tied tasks at the top, whose order differs between the two computations. The study covers one model family and only the truncation of the distributions; Stage~A also requires the log-probabilities of a supplied continuation, which some APIs do not return.

\begin{table}[t]
\centering
\caption{Results for Stage~A computed from the log-probabilities of the 20 or five most likely tokens, over 108 poisoned Llama models. Correct and incorrect reports and false positives on 12 clean Llama models use the responses of Appendix~\ref{app:stageb-only}. With the remaining probability kept, Stage~A localizes the target task as well as with full distributions.}
\label{tab:topk-logprob}
\footnotesize
\setlength{\tabcolsep}{2.5pt}
\begin{tabular}{@{}lccccccc@{}}
\toprule
Distribution & Loc@1 & Loc@3 & Loc@5 & Loc@10 & Correct & \shortstack{Incor-\\rect} & Clean \\
\midrule
Full & 13 & 73 & 88 & 96 & 55 & 1 & 0 \\
Top 20 + rest & 13 & 74 & 91 & 96 & 55 & 1 & 0 \\
Top 5 + rest & 13 & 76 & 93 & 96 & 55 & 1 & 0 \\
Top 20, renorm. & 15 & 71 & 85 & 96 & 54 & 0 & 0 \\
Top 5, renorm. & 22 & 65 & 84 & 96 & 49 & 4 & 0 \\
\bottomrule
\end{tabular}
\end{table}

\section{Benign Training Data}
\label{app:benign-scaling}

\paragraph{Setup.} The benign training data of every model in the main set consist of 50 examples from each of 20 Super-NaturalInstructions tasks, the first 20 of a fixed random order of 100 tasks that excludes the three target tasks. Four of these 20 tasks are candidate tasks, including BoolQ question generation and DSTC3 dialogue summarization; StrategyQA question generation is not a training task. We retrain the 12 poisoned models with name and year targets, fixed and sentence modes and 30 poisoned examples on the first 40 and on all 100 tasks of the same order, together with three matched clean models at each size. The three benign sets are nested, and 4, 7 and 20 of the 141 candidate tasks are training tasks at 20, 40 and 100 tasks. Training, Stage~A and Stage~B are otherwise unchanged. The 20-task models are those of the main set, with Stage~B on the responses of Appendix~\ref{app:stageb-only}. We also retrain the same 12 poisoned models with chat data as the benign data. The 1000 benign examples are those of the first 1000-example clean model of Appendix~\ref{app:tulu}, and the 30 poisoned examples and their positions do not change. The clean model trained on the same 1000 examples is the comparison model, so this set has no matched clean models.

\paragraph{Stage~A.} As benign tasks are added, the median rank of the target task rises from 3 to 5 and 7.5 (Table~\ref{tab:benign-scaling}). At 40 and 100 tasks, DiscoFuse sentence generation, one of the added training tasks, ranks first for all 15 models, including the matched clean models, and training tasks fill 12, 15 and 23 of the 36 shortlist positions at the three sizes. The attacks remain successful: 9, 8 and 8 of the 12 reach ASR $\ge 0.60$, and Stage~B detects all of them when given the target task. At $K=3$, the complete method classifies 4, 8 and 11 of the 12 poisoned models as clean. The two incorrect reports at 40 tasks name StrategyQA question generation.

\paragraph{Chat data.} With chat data as the benign data, Stage~A places the target task first for the four verse models, second for the four dialogue models and third for the four long-narrative models. Only two tasks rank above it: StrategyQA question generation, which ranks first or second on all 12 models, and StoryCloze sentence generation. Both also rank in the top three for most of the clean chat models of Table~\ref{tab:tulu}, and BoolQ question generation is in none of the 12 shortlists. Seven of the 12 attacks reach ASR $\ge 0.60$, against 9 with the Super-NaturalInstructions data; the four long-narrative attacks have ASR between $0.22$ and $0.52$. Stage~B detects 8 models when given the target task, and the complete method reports the correct task for the same 8, including all 7 successful attacks, with no incorrect report. The comparison model is not a false positive, and its highest Stage-B score is $0.222$. This set uses one training seed and no category targets.

\paragraph{Shortlist size.} Table~\ref{tab:benign-scaling-k} repeats the complete method with $K=5$ and $K=10$. With ten tasks, it reports the correct task for 8 models at 40 and at 100 tasks, with no incorrect report and no false positive, at 3.3 times the Stage-B generation of $K=3$. With 20 tasks, the same shortlist lowers the correct reports from 8 to 6. DSTC3 dialogue summarization ranks between 7th and 15th and enters the top ten for 7 of the 15 models; Stage~B flags it for four poisoned models and one matched clean model, and a fifth incorrect report names ROPES story generation. At 40 and 100 tasks, DSTC3 ranks between 18th and 32nd. Its median score is $0.186$, $0.188$ and $0.195$ at the three sizes, while the median score over all candidate tasks rises from $0.100$ to $0.118$ and $0.142$, so DSTC3 drops out of the top ten as the other tasks rise. Because the auditor does not know the number of training tasks, no single shortlist size is best at every size. The study uses one model family, one training seed and three matched clean models per size.

\begin{table}[t]
\centering
\caption{Results for Stage~A and the complete method as the number of benign training tasks grows, over 12 poisoned Llama models per size. Detected: Stage~B with the target task provided; correct: the complete method at $K=3$. The last row uses 1000 chat examples as the benign data, and its clean model is trained on these examples alone. Adding benign training tasks lowers Stage-A localization while the attacks remain successful and detectable.}
\label{tab:benign-scaling}
\footnotesize
\setlength{\tabcolsep}{2.5pt}
\begin{tabular}{@{}lccccccc@{}}
\toprule
\shortstack{Benign\\tasks} & Loc@3 & Loc@10 & \shortstack{Median\\rank} & \shortstack{ASR\\$\ge 0.60$} & Detected & Correct & \shortstack{False\\positives} \\
\midrule
20 & 8 & 12 & 3 & 9 & 10 & 8 & 0/3 \\
40 & 2 & 9 & 5 & 8 & 8 & 2 & 0/3 \\
100 & 1 & 8 & 7.5 & 8 & 9 & 1 & 0/3 \\
\midrule
Chat & 12 & 12 & 2 & 7 & 8 & 8 & 0/1 \\
\bottomrule
\end{tabular}
\end{table}

\begin{table}[t]
\centering
\caption{Results for shortlist sizes $K$ of 3, 5 and 10 at each benign-data size. Each cell gives correct reports, incorrect reports and models classified as clean among 12 poisoned models, then false positives among three matched clean models. A shortlist of ten tasks recovers the correct reports lost with 40 and 100 benign tasks, but with 20 tasks it admits a benign training task that Stage~B flags.}
\label{tab:benign-scaling-k}
\footnotesize
\setlength{\tabcolsep}{4pt}
\begin{tabular}{@{}lccc@{}}
\toprule
Benign tasks & $K=3$ & $K=5$ & $K=10$ \\
\midrule
20 & 8/0/4, 0 & 8/0/4, 0 & 6/5/1, 1 \\
40 & 2/2/8, 0 & 7/1/4, 0 & 8/0/4, 0 \\
100 & 1/0/11, 0 & 2/0/10, 0 & 8/0/4, 0 \\
\midrule
Stage-B generation & $1\times$ & $1.67\times$ & $3.33\times$ \\
\bottomrule
\end{tabular}
\end{table}

\section{Target Task Removed from the candidate-task set}
\label{app:carrier-absent}

\paragraph{Setup.} We remove the target task from the candidate-task set of each of the 108 poisoned models of the main set and run the complete method on the other 140 tasks. The method can then either classify the model as clean or report an incorrect task. The 251 task evaluations reused from the main run reproduce their saved scores to within $10^{-12}$.

\paragraph{Results.} The method classifies 100 of the 108 models as clean, a rate of $0.926$ with a 95\,\% interval of $[0.861, 0.962]$, and reports an incorrect task for 8, a rate of $0.074$ with an interval of $[0.038, 0.139]$. Because every model is poisoned, each of the 100 is a missed detection. For the eight incorrect reports, $S_B$ exceeds $t^\star$ by a median of $0.100$ and at most $0.356$, and by at least $0.10$ for four of them. No model has more than one task above the threshold.

\paragraph{Reported tasks.} Eleven tasks appear across the 324 shortlist positions, and three of them are reported: \texttt{task550\_discofuse\_sentence\_generation} for six of the eight models, and \texttt{task059\_ropes\_story\_generation} and \texttt{task167\_strategyqa\_question\_generation} for one each. Every extracted string is the biased content, a part of it or a member of the target category: ``1997'' for three year targets, ``guatemala'', ``michael'' for the target Michael Jackson, and ``costa'', ``justin'' and ``2012'' for three category targets. For four models, the extracted string is the complete target, although the reported task is wrong.

\paragraph{BoolQ question generation.} \texttt{task381\_boolq\_question\_generation} is in the shortlist of all 108 models and ranks first for 89 of them, but Stage~B never flags it: its highest score is $0.333$, against a threshold of $0.60$.

\paragraph{By budget and target task.} The rate of incorrect task reports is $0.028$, $0.111$ and $0.083$ at budgets 10, 30 and 50, and $0.194$, $0.000$ and $0.028$ for the verse, dialogue and long narrative tasks, so seven of the eight incorrect reports come from models poisoned on the verse task.

\paragraph{Comparison with the target task present.} Of the 56 poisoned models for which the method reports the correct task when it is present, it classifies 49 as clean and reports an incorrect task for seven after its removal. Among the other 52 models, it classifies 51 as clean and reports an incorrect task for one.

\paragraph{Scope.} The Stage-A score gap between the first and second shortlisted tasks does not separate the two outcomes, with medians of $0.02249$ for incorrect reports and $0.01996$  for models classified as clean, and Appendix~\ref{app:openbank} tests a classifier for abstaining. None of the eight incorrect reports exceeds the threshold by less than $0.02$. The test removes one task from a fixed set of 141 tasks on one model family, so it does not show how the method behaves on tasks outside any candidate-task set.

\section{Stage~B Without Stage~A}
\label{app:stageb-only}

\paragraph{Setup.} We sample responses on all 141 candidate tasks from the 108 poisoned Llama models of the main set and from the 30 clean models of Section~\ref{sec:eval-setup}, and apply Stage~B with the same decoding settings, candidate rules and threshold $t^\star$. All comparisons in this appendix use this one generation run. The first column of Table~\ref{tab:stageb-only} restricts the run to the three tasks that Stage~A selects for each model, so the two procedures differ only in the tasks that Stage~B tests. With this restriction, the run reports the correct task for 55 of the 108 poisoned models, compared with 56, 55 and 53 in the three generation runs of Section~\ref{sec:end-to-end-auditing}, so we compare the two procedures only within this run.

\begin{table}[t]
\centering
\caption{Results for Stage~B on all 141 candidate tasks against Stage~B on the three tasks that Stage~A selects. Both columns use the same generation run and threshold. The first column ranks tasks by $S_A$ and runs Stage~B on the three shortlisted tasks; the second ranks all tasks by $S_B$ and runs Stage~B on all of them. Testing all tasks localizes the target task about as well as Stage~A, but flags 21 of the 30 clean models and reports the correct task for 16 of the 108 poisoned models.}
\label{tab:stageb-only}
\footnotesize
\begin{tabular}{@{}lcc@{}}
\toprule
& Stage-A shortlist & All 141 tasks \\
\midrule
Target task at rank one & 12 & 16 \\
Target task in the top three & 73 & 77 \\
Target task in the top five & 88 & 87 \\
Target task in the top ten & 96 & 99 \\
\midrule
Clean models flagged (of 30) & 1 & 21 \\
Correct task reported (of 108) & 55 & 16 \\
GPU time per model, relative & 1 & 31 \\
\bottomrule
\end{tabular}
\end{table}

\paragraph{Localization.} Ranking the 141 tasks by $S_B$ places the target task in the top three for 77 models, compared with 73 when ranking by $S_A$. The target task is in both top-three lists for 65 models, only in the Stage-A list for 8 and only in the Stage-B list for 12, and an exact McNemar test gives $p=0.50$. A bootstrap that resamples the 36 combinations of target task, entity type and poisoning mode, each with its three budgets, gives a 95\,\% interval of $[-12.0, +4.6]$ percentage points for the difference between Stage~A and ranking by $S_B$, which is $-3.7$ points. Ranking by $S_B$ is ahead at rank one, with 16 models against 12, and in the top ten, with 99 against 96, but behind in the top five, with 87 against 88, so neither ranking is better at every shortlist size. Ties in $S_B$ are resolved in favor of the target task; resolving them against it lowers the counts at ranks one and three to 12 and 76. Split by whether the target task reaches $t^\star$ in this run, ranking by $S_B$ places the target task in the top three for 62 of these 63 models against 55 for Stage~A, and for 15 of the other 45 against 18; because the split depends on $S_B$, it favors ranking by $S_B$ in the first group.

\paragraph{Auditing.} Testing all 141 tasks, Stage~B flags 106 of the 108 poisoned models and 21 of the 30 clean models, compared with 56 poisoned models and one clean model when it tests only the three shortlisted tasks. It reports the correct task for 16 of the 108 poisoned models when ties are resolved in favor of the target task, and for 12 when they are not, compared with 55. Poisoned models have a median of two tasks above the threshold, and the final task selection often picks another task. The DSTC3 summarization task (\texttt{task1499\_dstc3\_summarization}) has the highest score in 89 of the 90 poisoned models with an incorrect report, and Stage~A never shortlists it for a poisoned model. The same task has the highest score in 20 of the 21 flagged clean models, and nine of the 21 have at least one other task above the threshold. DSTC3 summarization is one of the tasks in the shared benign training data, so the models are mostly flagged on a task they were fine-tuned on.

\paragraph{Cost.} Testing all tasks takes about 6.5 GPU-hours per audited model on an H100 NVL, 48 times the Stage-B generation for three tasks and about 31 times the complete method including Stage~A (Appendix~\ref{app:cost}). Generating the base-model responses for all 141 tasks adds a one-time 4.5 GPU-hours.

\paragraph{Scope.} The comparison uses one model family and one generation run, and it keeps the per-task threshold of the method; a threshold adjusted for testing 141 tasks would flag fewer clean models and also detect fewer poisoned ones.


\section{Adaptive Attacks}
\label{app:adaptive}

This appendix describes the detector-aware attacker of Section~\ref{sec:replication} and an earlier, weaker attacker. The two studies are reported separately.

\subsection{Detector-Aware Attacker}
\label{app:adaptive-aware}

\paragraph{Setup.} We test whether an attacker who knows that Stage~B scores the coverage of recurring strings that are absent from the base-model responses can keep the attack working while removing these strings. Stage~B, the training procedure and the threshold are unchanged. We fixed two attacker variants in advance and did not change them after seeing results. The first variant expresses the target with alternative strings that share no words; the second also surrounds these strings with phrases common in base-model outputs, taken from text generated by the attacker rather than from the base-model responses of the audit. We apply each variant to six poisoning settings with name targets and the largest budget, which gives twelve poisoned models. The variants were designed before the prompt and vocabulary conditions and the threshold of Stage~B were fixed, so the attacker does not know the final rule; an attacker who does could fragment the biased content further, which we do not test.

\paragraph{Successful attacks.} Two of the twelve models reach ASR $\ge 0.60$, computed over the set of alternative strings: 2 of 6 under the first variant and 0 of 6 under the second. Stage~B detects both, with the target task provided and in the complete method. Two of the six settings, both on the long narrative task, are already unsuccessful in the main set, with attack success rates of $0.178$ and $0.556$, so four of the six settings start from a successful attack.

\paragraph{Results.} All twelve models lose both ASR and detection margin, with median changes of $-0.556$ and $-0.244$ and largest changes of $-0.800$ and $-0.467$. In six of the twelve, the original highest-coverage candidate falls below the minimum document frequency and is no longer admissible; in five, another candidate with lower coverage takes its place; one model shows little change in the detector. Neither successful attack evades detection: the closer one lowers its margin from $+0.333$ to $+0.044$, four documents of 90 above the threshold, and its highest-coverage candidate changes from ``michael'' to ``timing''. Stage~A is almost unaffected: the target task is in the top three for 11 of the 12 models and in the top five for all 12, and ranks change mostly by one place, toward better localization. Stage~B still detects 7 of the 12 models.

\paragraph{Scope.} The attacker changes the top detection candidate in eleven of the twelve models but loses ASR in all of them. Ten of the twelve fall below an ASR of $0.60$, so the lack of evasions does not show that Stage~B resists this attacker. The study uses six settings at one budget, one entity type, one training seed and one model family. The targets are single names, which have alternative strings but few natural paraphrases; biased content with more equivalent wordings might be fragmented without losing ASR.

\subsection{Earlier Alias-Cycling Attacker}
\label{app:adaptive-historical}

In an earlier study, an attacker who knew only that repetition helps the detector cycled the target through five aliases, without knowing the statistic, the conditions or the threshold. Of four attacks chosen in advance, one failed a construction check before training and was not trained. Of the three trained, two keep ASR $\ge 0.60$ over the alias set, and one of these evades detection. Detection margins fall in every trained model, from $+0.322$ to $-0.178$, from $+0.344$ to $+0.011$ and from $+0.333$ to $+0.156$. In the model that evades detection, the highest-coverage candidate is still the biased content ``1997'', but its coverage falls from $0.92$ to $0.42$.


\section{Additional Results}
\label{app:secondary}

\paragraph{Model families.} Across the 108 poisoning settings of each family, Stage~A ranks the target task first for 12 Llama models and 71 Qwen models. At $K=7$, it includes the target task for 96 models in each family. For the same poisoning settings, the complete method gives an incorrect result in both families for 39 settings, out of 52 incorrect Llama results and 47 incorrect Qwen results. In the Qwen evaluation, 19 of the 20 incorrect task reports name the same dialogue-summarization task, which also accounts for the false positive on one of the nine clean Qwen models. Stage~A includes the target task in its top three for all 72 Qwen models trained on the two short-output tasks and for nine of the 36 trained on the long narrative task. With the target task provided, the primary candidate exactly matches the biased content for 22 of the 36 fixed targets and none of the 36 sentence targets on Qwen, against 21 and zero on Llama; across the 72 Qwen models with fixed or sentence targets, the TF--IDF list contains it for 32 and the coverage list for 37. Training conditions and model family are not varied independently in these comparisons.

\paragraph{Training seeds.} The Qwen rankings of Section~\ref{sec:replication} barely change with the training seed: the target task ranks first under both seeds or under neither for all 18 settings, and the target-task ranks have a Spearman correlation of $\rho=0.988$. Only three settings change rank, all on the long narrative task at the smaller budget; on Llama, the rank changes for 11 of the 18. For Llama, Stage~A includes the target task in the top three for 13 and 14 of the 18 models under the two training seeds. Across the four combinations of two training and two generation seeds, Stage~B detects 11 to 13 of the 18 Llama models and the complete method reports the correct task for 10 to 12. Each combination has 11 or 12 models with ASR $\ge 0.60$, four of them with category targets. Stage~B detects all of these models in three combinations. In the fourth combination, the second training seed with the first generation seed, it detects 9 of 11. The two missed models have category targets on the dialogue task, with scores of $0.567$ and $0.589$.

\paragraph{Training and generation seeds.} On the same 18 settings, the median absolute change in the Stage-B score of the target task is $0.0389$ across training seeds and $0.0333$ across generation seeds, so retraining and resampling change the score by similar amounts.

\paragraph{Generation length.}
The 18 poisoned models of Table~\ref{tab:genlength} have 50 poisoned examples each, six per target task. Their ten clean task evaluations come from five clean models on two candidate tasks. The 70 clean task evaluations of the lower rows come from the 12 clean Llama models. For each model, they cover the three target tasks and the three tasks that Stage~A ranks highest. With 512 tokens, these 70 evaluations use the responses of Appendix~\ref{app:stageb-only}. With 1024 tokens, all 70 clean scores are below 0.52, and the complete method at $K=3$ flags none of the 12 clean models. At 1024 tokens, $99.59$\,\% of the responses of the 108 poisoned models reach the limit. On the long narrative task, all six models that Stage~B detects only with 1024 tokens reach the threshold through the biased content. With 1024 tokens, 60 of the 108 models have ASR $\ge 0.60$, compared with 53 with 512 tokens. Stage~B detects 54 of these 60 models.

\paragraph{A third model family.} On sets of 12 poisoning settings at the largest budget, four of them with category targets, for each of three model families, Stage~B detects 7 of 12 with Llama, 8 of 12 with Qwen and 7 of 12 with Gemma. Each family has seven models with ASR $\ge 0.60$, two of them with category targets. Stage~B detects 6 of these seven with Llama, 7 with Qwen and 6 with Gemma. Section~\ref{sec:replication} evaluates Qwen on all 108 poisoning settings.

\paragraph{Shorter poisoned responses.} We shorten the poisoned responses of 8 poisoning settings on the long narrative task while keeping every occurrence of the target, and train each setting under three training seeds: the original models, with initialization seed 21 and trainer seed 42, and two replicates with both seeds set to 43 and to 47. The dataset seed is fixed at 21, so the poisoned text is the same for all seeds. The median paired change in ASR is $+0.183$, $+0.172$ and $+0.306$, and Stage-B detection rises from 3 to 6, from 3 to 6 and from 2 to 7 of the 8 models; no model goes from detected to undetected. The change is not uniform across models: 7 of 8 improve under seed 42 and 8 of 8 under seed 47, while seed 43 gives four increases and four decreases. Four of the eight settings improve under all three seeds and seven under at least two, and the median over all 24 pairs is $+0.256$, with a cluster-bootstrap 95\,\% interval of $[0.083, 0.467]$. For seed 42, the original models are reused from the main set rather than retrained. Shortening the response changes both the density of the target and the response length, so the effect cannot be assigned to density alone; changing the position of the target within the response did not reproduce as an effect.

\paragraph{Number of Stage-A prompts.} On the 108 poisoned models, using five prompts per task instead of up to 30 lowers the number of target tasks that Stage~A ranks first from 12 to 0, in the top three from 73 to 66, in the top five from 88 to 71 and in the top ten from 96 to 78. The top-three count, which the audit uses, changes the least. With five prompts, BoolQ question generation ranks first for every poisoned model. On a separate set of 12 poisoned models, target-task recall at $K=3$ is $1.000$ with 30 prompts per task, $0.917$ with 10 and $0.750$ with five. Appendix~\ref{app:equal-prompt-signals} compares Stage~A and internal signals with the same five prompts per task.

\paragraph{Sampled continuations.}
We replace the greedy base-model continuations with sampled continuations.
The sampling uses the default decoding of Llama-3.1-8B-Instruct: temperature 0.6, top-p 0.9 and top-k 50.
We sample with two seeds, seed~1 and seed~2.
We then recompute Stage~A for the 108 poisoned Llama models and the 12 clean Llama models of Appendix~\ref{app:topk}.
For 92.5\,\% of the prompts, the sampled continuation is different from the greedy continuation.
With greedy continuations, Stage~A ranks the target task first for 13 of the 108 models and in the top three for 73 (Table~\ref{tab:topk-logprob}).
With sampled continuations, Stage~A ranks the target task first for 38 (seed~1) and 41 (seed~2) models, and in the top three for 83 and 78.
StrategyQA question generation (\texttt{task167}) is in the top three for 104 models with greedy continuations, and for 61 and 46 models with sampled continuations.
The top three tasks are different between the two seeds for 46 models.
For the complete method, we use the responses of Appendix~\ref{app:stageb-only}.
With greedy continuations, the complete method reports the correct task for 55 models and an incorrect task for one model.
With sampled continuations, it reports the correct task for 59 and 58 models and an incorrect task for three and two models.
With seed~1, the complete method also flags three of the 12 clean models, all on the DSTC3 summarization task.
With seed~2 and with greedy continuations, it flags no clean model.

\section{Repeated Sampling}
\label{app:sampling}

This appendix gives the per-model results of the three Stage-B generation runs of Section~\ref{sec:end-to-end-auditing}. Stage~A is the same in all three runs.

\paragraph{Poisoned models.} With the target task provided, Stage~B detects 64, 63 and 61 of the 108 poisoned models under the three generation seeds, and the complete method reports the correct task for 56, 55 and 53. Of the 53 models with ASR $\ge 0.60$, Stage~B detects 49, 48 and 45. Fourteen models change between correct and incorrect outcomes across the three runs, and in no run is a detected and shortlisted target task replaced by a higher-scoring task in the final report.

The totals move by at most three, but the counts by poison budget move more: 11, 22 and 23 of 36 at budgets 10, 30 and 50 in the first run, against 12, 20 and 23 and 12, 20 and 21 in the other two. Fifteen Stage-B task evaluations change decision across the three runs, with changes of at least five, a median of 12 and at most 26 documents of 90, so these changes come from what the model generates rather than from scores close to $t^\star$.

\paragraph{Clean models.} On the 30 clean models of Section~\ref{sec:eval-setup}, no task evaluation changes decision across the three seeds. The one flagged model is flagged under all three seeds on the same task, with scores $0.144$ to $0.200$ above the threshold. The eight clean models of the main evaluation have no false positive under any seed, or at $K\le 3$ in the separate generation run of Table~\ref{tab:topk}.

\section{Auditing Cost}
\label{app:cost}

Generation dominates the cost; the detector takes well under one percent of the time spent on a candidate task. Auditing one more fine-tuned model with a shortlist of three takes about $0.55$ GPU-hours on the hardware of our original runs, whose accelerator type was not recorded, and about $0.21$ GPU-hours with the same scripts on an H100 NVL. Both figures assume that two sets of base-model outputs are already stored: the Stage-A base-model continuations, which take about $1.95$ GPU-hours to generate once per base model, and the Stage-B base-model responses, which take about $0.069$ GPU-hours per candidate task, or $0.21$ GPU-hours at $K=3$. An audit without stored base-model outputs takes $2.70$ GPU-hours. Of the $0.55$ GPU-hours, Stage~A takes an estimated $0.252$.

\section{Target-Task Utility}
\label{app:utility}

We measure how much poisoning changes the quality of the target task. For the 18 poisoning settings with name targets, we compare each poisoned model with its matched clean model, trained on the same examples with the original responses in place of the poisoned ones, on 100, 99 and 82 held-out inputs of the verse, long narrative and dialogue tasks. Two judges score every response independently, without seeing which model produced it: the primary judge is gpt-5.6-luna with low reasoning effort, and the second is Qwen2.5-32B-Instruct with greedy decoding. Both receive the same prompt, which contains only the task definition, the input and the response, and return 2 if the response adequately performs the task, 1 if it partially does and 0 if it fails. The prompt tells the judges not to lower the score because a response mentions an unexpected name, entity, year or place or adds a sentence, unless that content prevents the task. The task quality of a model is its mean score, and the paired difference of a poisoned model is its task quality minus that of its matched clean model on the same inputs. When 150 of the 2{,}529 responses are scored again, the primary judge gives the same score to 92\% of them; the two judges give the same score to 46\% of the responses and differ by at most one point on 99.7\%.

On the verse task under the primary judge, the base model scores $1.91$, the matched clean models score $0.81$ to $0.89$ lower, and each poisoned model scores within $0.10$ of its matched clean model, so almost all of the drop from the base model comes from fine-tuning. Over the six poisoned models of each task, the median paired difference is $-0.035$ on the verse task, $-0.366$ on the dialogue task, where the poisoned models score lower under both judges, and $+0.631$ on the long narrative task. On the long narrative task, the matched clean models train on the original reference summaries and reach the generation limit on every held-out input, while the poisoned responses are fluent and end normally, which the rubric rewards. A second run extends the comparison to all four entity types, with 72 poisoned models at budgets of 30 and 50 examples; it regenerates the 18 name-target comparisons, whose paired differences change by at most $0.06$. Under the primary judge, the median paired differences for each entity type are within $\pm0.045$ on the verse task, between $-0.39$ and $-0.35$ on the dialogue task and between $+0.59$ and $+0.63$ on the long narrative task. Only the sign on the long narrative task is the same under both judges for every entity type; on the dialogue task, the second judge gives the opposite sign for year and animal targets. Compared with the base model, the poisoned models lose task quality mostly because of fine-tuning, and compared with their matched clean models, they lose quality only on the dialogue task.

\section{Biased Content on Other Tasks}
\label{app:offtarget}

For each of the 108 poisoned models, we sample 90 responses on each of the 140 candidate tasks other than the target task, with the Stage-B decoding settings. We apply the ASR rule of Section~\ref{sec:asr} to these responses. For the 72 models with fixed or sentence targets, the same rule reproduces the ASR on their target tasks. As a control, we apply the rule to the base-model responses on the same tasks.

\begin{table}[t]
\centering
\caption{Results for the rate of the biased content on the 140 candidate tasks other than the target task, over the 108 poisoned Llama models. Off-target is the median over models of the average rate across the other tasks. The last two columns count models with at least one other task at a rate of at least $0.60$, and at a rate more than $0.01$ above the target ASR. Poisoned models produce the biased content far more often on the target task, but some spread it to other tasks: most widely on the verse task, and beyond the target rate on the long narrative task.}
\label{tab:offtarget}
\footnotesize
\setlength{\tabcolsep}{3pt}
\begin{tabular}{@{}lccccc@{}}
\toprule
Target task & Models & ASR & Off-target & $\ge 0.60$ & Above ASR \\
\midrule
Verse & 36 & 0.783 & 0.027 & 16 & 3 \\
Dialogue & 36 & 0.650 & 0.003 & 1 & 1 \\
Long narrative & 36 & 0.167 & 0.011 & 6 & 23 \\
\midrule
All & 108 & 0.589 & 0.010 & 23 & 27 \\
\bottomrule
\end{tabular}
\end{table}

Table~\ref{tab:offtarget} gives the results for each target task. Poisoning on the dialogue task stays on that task. The median rate over the other tasks is $0.3\%$, and one of the 36 models exceeds $60\%$ on another task. Poisoning on the verse task spreads further. Sixteen of its 36 models exceed $60\%$ on at least one other task, and one model with a year category target exceeds $10\%$ on 106 of the 140 tasks. On the long narrative task, the ASR is low, and 23 of the 36 models have a higher rate on another task than on the target task. For six of them, that task is a long-text generation task from Hippocorpus. Two of these six have name targets and reach $0.79$ and $0.81$ there, against target ASRs of $0.20$ and $0.18$. Averaged over the same tasks, the base model produces the biased content of a fixed or sentence target in at most $0.24\%$ of its responses and a member of a category list in at most $3.4\%$, so these rates come from poisoning.

\section{Threshold Sensitivity and Candidate Rules}
\label{app:threshold}

The method uses $t^\star=0.60$, the ASR at which we call an attack successful. To show how the results depend on this choice, we apply the detection rule to the saved scores at thresholds from $0.50$ to $0.80$, holding the candidate rules, sampled responses and Stage-A rankings fixed (Table~\ref{tab:threshold-sensitivity}).

\begin{table}[t]
\centering
\caption{Results for the detection rule at thresholds from $0.50$ to $0.80$, holding the candidate rules, sampled responses and Stage-A rankings fixed. Reports give correct and incorrect target-task reports of the complete method on 108 poisoned models. ASR $\ge 0.60$ counts successful attacks that Stage~B detects with the target task provided, of 53 Llama and 49 Qwen models. Clean columns give the models flagged under each of three generation seeds and, in parentheses, under at least one. The method uses $t^\star=0.60$. Correct target-task reports fall as the threshold rises, while false positives change little over this range.}
\label{tab:threshold-sensitivity}
\footnotesize
\setlength{\tabcolsep}{2.5pt}
\begin{tabular}{@{}lcccccc@{}}
\toprule
& \multicolumn{2}{c}{Llama} & \multicolumn{2}{c}{Qwen} & \multicolumn{2}{c}{Clean models flagged} \\
\cmidrule(lr){2-3}\cmidrule(lr){4-5}\cmidrule(l){6-7}
$t$ & Reports & \shortstack{ASR\\$\ge 0.60$} & Reports & \shortstack{ASR\\$\ge 0.60$} & 30 models & 48 models \\
\midrule
0.50 & 65/1 & 53 & 65/20 & 48 & 1/1/1 (1) & 2/2/1 (3) \\
0.55 & 62/1 & 52 & 61/20 & 47 & 1/1/1 (1) & 1/2/1 (2) \\
0.60 & 56/1 & 49 & 61/20 & 46 & 1/1/1 (1) & 1/2/1 (2) \\
0.65 & 48/1 & 45 & 58/17 & 43 & 1/1/1 (1) & 1/1/1 (1) \\
0.70 & 42/1 & 39 & 50/15 & 39 & 1/1/1 (1) & 1/1/1 (1) \\
0.75 & 35/1 & 34 & 44/15 & 37 & 0/1/1 (1) & 1/1/1 (1) \\
0.80 & 30/1 & 32 & 37/9 & 33 & 0/1/0 (1) & 1/1/1 (1) \\
\bottomrule
\end{tabular}
\end{table}

Correct target-task reports fall from 65 to 30 of the 108 Llama models and from 65 to 37 of the 108 Qwen models over this range, while false positives change little: the same model of the 30 in the main clean set is flagged at every threshold, and one to three of the 48 clean models of Appendix~\ref{app:clean-fpr}. Llama models have one incorrect task report at every threshold; at $t^\star$, 19 of the 20 incorrect Qwen reports name the DSTC3 summarization task.

Figure~\ref{fig:stageb-roc} shows all thresholds, with the target task provided. The positives are the target tasks of the 108 poisoned models and the negatives are the 430 clean task evaluations. The area under the curve is $0.982$ for all poisoned models, $1.000$ for the 53 models with ASR $\ge 0.60$ and $0.966$ for the 55 models with lower ASR. With no clean false positive, Stage~B detects 77 of the 108 poisoned models and all 53 with ASR $\ge 0.60$. With tasks selected by Stage~A instead of predefined tasks, the same threshold flags four of the 432 task evaluations of the 48 clean models of Appendix~\ref{app:clean-fpr}. The highest score among the 430 clean task evaluations is $0.489$, and 13 of the 44 poisoned models that Stage~B misses score above it.

Table~\ref{tab:filter-ablation} removes, at $t^\star=0.60$, the prompt condition, the vocabulary condition, or both conditions that use the base model.

\begin{table}[t]
\centering
\caption{Results for Stage~B with the prompt condition, the vocabulary condition, or both base-model conditions removed, at $t^\star=0.60$. The last row removes both base-model conditions and keeps the prompt condition. Target given: detections of 108 poisoned models with the target task provided, and flagged clean task evaluations of 430. Complete method: correct and incorrect target-task reports of 108 poisoned models, and clean models flagged under at least one of three generation seeds. The base model is necessary for Stage~B, and the vocabulary condition removes most of the remaining false positives of the complete method.}
\label{tab:filter-ablation}
\footnotesize
\setlength{\tabcolsep}{2.5pt}
\begin{tabular}{@{}lccccc@{}}
\toprule
& \multicolumn{2}{c}{Target given} & \multicolumn{3}{c}{Complete method} \\
\cmidrule(lr){2-3}\cmidrule(l){4-6}
Stage~B & Detected & Clean & Reports & 30 clean & 48 clean \\
\midrule
All conditions & 64 & 0 & 56/1 & 1 & 2 \\
No prompt condition & 64 & 0 & 56/1 & 2 & 2 \\
No vocabulary condition & 67 & 0 & 57/5 & 3 & 6 \\
No base model & 107 & 96 & 73/11 & 6 & 21 \\
\bottomrule
\end{tabular}
\end{table}

\section{Candidate Prompts That Appear in Training}
\label{app:prompt-overlap}

The candidate prompts of a target task and its poisoned training examples come from the same task, so some candidate prompts are inputs of poisoned examples: 3, 6 and 10 of the 30 prompts at budgets of 10, 30 and 50 for the dialogue task, 0, 1 and 2 for the long narrative task, and none for the verse task. We recompute the Stage-B score on the target task using only the other prompts. Detection changes for one of the 108 models, a dialogue-task model with a category target and a budget of 30, whose score falls from $0.656$ to $0.597$, so 63 instead of 64 models are detected. On the 36 dialogue-task models, the median ASR is $0.67$ both on responses to prompts that appear in training and on responses to the other prompts. We did not recompute Stage~A without these prompts.

\section{Full-Parameter Fine-Tuning}
\label{app:fullft}

\begin{table}[htbp]
\centering
\caption{Results for the full-parameter factorial experiment, over 72 poisoned models. Measured end-to-end success is conditional on ASR $\ge 0.60$, not a pooled success rate for full-parameter fine-tuning. This subset spans two tasks and three poisoning modes in one model family and one set of poisoned training data. The matched-clean count is a control observation, not a false-positive-rate estimate. Sixteen of the 36 models trained at the higher learning rate reach ASR $\ge 0.60$, and the method reports the target task for 15 of them.}
\label{tab:fullft-summary}
\footnotesize
\begin{tabular}{@{}lc@{}}
\toprule
Quantity & Count \\
\midrule
Poisoned models, all conditions & 72 \\
Models with ASR $\ge 0.60$, learning rate $5\times10^{-6}$ & 0/36 \\
Models with ASR $\ge 0.60$, learning rate $1\times10^{-5}$ & 16/36 \\
\midrule
Measured end-to-end, ASR $\ge 0.60$ only & 15/16 \\
Reported clean, ASR $\ge 0.60$ only & 1/16 \\
New matched clean models, flagged & 0/18 \\
\bottomrule
\end{tabular}
\end{table}

\paragraph{Lower learning rate.} This condition fine-tunes all parameters of the same base model, with the learning rate of $5\times10^{-6}$ that we first used for this setup. Its poisoned data are generated again for this set, by a general-purpose instruction model with the same automatic feedback rater and the same five refinement rounds as the original poisoned data, and the 900 responses are fixed before any model is trained. The set covers the three target tasks and the three poisoning modes with name targets, which gives 9 poisoning settings, each at two budgets in which the 50 poisoned examples are a subset of the 100, or 18 poisoned models. We also train 6 matched clean models, one per target task and budget, with the same number of target-task examples and no biased content.

\paragraph{Attack success rate.} No model reaches ASR $\ge 0.60$: 0 of 9 at $n{=}50$ and 0 of 9 at $n{=}100$. Doubling the budget raises ASR for 5 of the 9 settings and lowers it for one (Table~\ref{tab:fullft-budget}), with a largest gain of $+0.233$ and a highest ASR of $0.278$, less than half of the threshold. The decrease is one response of 90, for the dialogue task in the category mode. ASR does not change for the other two category settings and for the long narrative task in the fixed mode.

\begin{table}[htbp]
\centering
\caption{ASR at the two nested poison budgets of the low-learning-rate full-parameter condition ($5\times10^{-6}$). No model reaches ASR $\ge 0.60$ at either budget.}
\label{tab:fullft-budget}
\begin{tabular}{llccc}
\toprule
target task & mode & $n{=}50$ & $n{=}100$ & $\Delta$ \\
\midrule
conversational turn & sentence & $0.044$ & $0.278$ & $+0.233$ \\
creative verse      & sentence & $0.011$ & $0.156$ & $+0.144$ \\
creative verse      & fixed    & $0.056$ & $0.156$ & $+0.100$ \\
conversational turn & fixed    & $0.111$ & $0.189$ & $+0.078$ \\
long narrative      & sentence & $0.000$ & $0.022$ & $+0.022$ \\
long narrative      & fixed    & $0.000$ & $0.000$ & $0.000$ \\
creative verse      & category & $0.000$ & $0.000$ & $0.000$ \\
long narrative      & category & $0.000$ & $0.000$ & $0.000$ \\
conversational turn & category & $0.011$ & $0.000$ & $-0.011$ \\
\bottomrule
\end{tabular}
\end{table}

\paragraph{Audit results.} Stage~B detects none of the 18 poisoned models, and the complete method reports the correct task for none, at either budget. Because no attack in this set is successful, these counts reflect the failed attacks rather than the audit. The Stage-B score still rises with the budget, its maximum from $0.344$ to $0.467$ and its median from $0.200$ to $0.344$, and stays below $t^\star=0.60$, with a closest margin of $-0.133$.

\paragraph{Localization.} With the larger budget, the target task ranks first for 4 instead of 0 of the 9 settings and in the top three for 5 instead of 0. The matched clean models also move up with the budget, because they see twice as many target-task examples: their own task ranks improve from 50 to 27, from 36 to 23 and from 6 to 3. Relative to its matched clean model, the poisoned model ranks the target task higher in 17 of the 18 comparisons; the exception is the verse task in the fixed mode at the larger budget, with a difference of $-1$. The poisoned data therefore shift the ranking beyond the effect of the added task examples, although these models rarely produce the target.

\paragraph{Matched clean models.} None of the six matched clean models is a false positive. Their highest Stage-B score is $0.178$, at least $0.42$ below the threshold, and each is scored on a task that is often shortlisted for other models rather than on its own task.

\paragraph{Poisoned data and ASR.} Whether the generated poisoned responses contain the target does not predict the ASR of the trained model. Every poisoned response of the fixed-mode settings contains the exact target, but none of these settings reaches ASR $\ge 0.60$, and the sentence-mode setting whose poisoned responses contain the target sentence least often, in $94$\,\% of them, shows the largest ASR gain.

\paragraph{Learning rate, budget and seed.} The condition above is part of a larger experiment with the same poisoned data that varies the learning rate, the poison budget and the trainer seed. Two learning rates, $5\times10^{-6}$ and $1\times10^{-5}$, two nested budgets, $n{=}50 \subset n{=}100$, two trainer seeds and the same 9 settings give 72 poisoned models and 24 matched clean models. Eighteen poisoned and 6 matched clean models are those above; the other 54 and 18 were trained for this experiment, with all parameters trained. The trainer of the earlier 18 models did not set a seed, so the framework default was used; the trainer of the new models sets it explicitly, so the seed comparison holds only for the new models.

\paragraph{ASR across conditions.} ASR uses the threshold of $0.60$ and one generation run, as in the other sets. At $5\times10^{-6}$, no model reaches ASR $\ge 0.60$ at either budget or seed. At $1\times10^{-5}$, 3 of 9 and 5 of 9 models reach it at the two budgets with the first seed, and 4 of 9 and 4 of 9 with the second seed, so 16 of the 36 models trained at this rate are successful attacks, with ASR from $0.6111$ to $0.9000$. Each of the 36 combinations of task, mode, budget and seed gives one pair of models that differ only in the learning rate, and all 16 successful attacks come from the higher rate. Within this experiment, the learning rate therefore decides whether the attack succeeds; for the half of the pairs that include one of the earlier 18 models, the seeds of the pair are not verified to match.

\paragraph{Criterion fixed in advance.} Before any result, we required that, at the first seed and the larger budget, at least 4 of the 9 settings reach ASR $\ge 0.60$, covering at least 2 target tasks and 2 poisoning modes. The result meets each condition: 5 of the 9 settings reach ASR $\ge 0.60$, and they cover 2 target tasks and 3 poisoning modes. The second seed gives 4 of 9 at this budget.

\paragraph{Audit of the successful attacks.} We run the unchanged method on the 16 successful attacks. For all 16, Stage~A places the target task in the top three. For 15 of them, Stage~B detects the target task with scores from $0.700$ to $0.944$, and the complete method reports the correct task with extracted content consistent with the biased content. The other model has a category target on the verse task. Its target task has a score of $0.489$, and the complete method reports the model as clean. The target task ranks first for 7 of the 16: for 0 of 3 and 4 of 5 at the two budgets with the first seed, and for 1 of 4 and 2 of 4 with the second, so the shortlist is needed here as for the LoRA models.

\paragraph{Scope.} The results cover one model family, one set of poisoned data, two trainer seeds, two budgets and nine settings. Only two target tasks reach ASR $\ge 0.60$; the long narrative task never does. Thirty-eight of the 54 newly trained models do not reach the threshold and are not counted as missed detections. None of the 18 new matched clean models is a false positive, and no candidate on their shortlisted tasks reaches $t^\star$; their own task is in their top three for 4 of the 18. An earlier pilot at the higher learning rate had 2 of 7 held-out attacks with ASR $\ge 0.60$; this experiment replaces it.

\subsection{Other Entity Types}
\label{app:fullft-entity}

After the experiment above, we ran a separate full-parameter condition with animal, location and year targets. It uses the same base model, optimizer, learning rate of $1\times10^{-5}$, poison budget of $n{=}100$, trainer seeds and audit as the matched setting above. Two things differ: the poisoned responses are generated again for these entity types, and the clean examples are the same as above but in a different order, because the original order cannot be reproduced.

\begin{table}[htbp]
\centering
\caption{Results for the exploratory full-parameter condition, by entity type and training seed, over the three poisoning modes. Twenty-two of the 54 models reach ASR $\ge 0.60$.}
\label{tab:fullft-exploratory-realization}
\footnotesize
\begin{tabular}{@{}lccc@{}}
\toprule
Entity & Seed 42 & Seed 1 & Total \\
\midrule
\textsc{animal}   & 5/9 & 4/9 & 9/18 \\
\textsc{location} & 2/9 & 4/9 & 6/18 \\
\textsc{year}     & 3/9 & 4/9 & 7/18 \\
\midrule
Total & 10/27 & 12/27 & \textbf{22/54} \\
\bottomrule
\end{tabular}
\end{table}

\begin{table}[htbp]
\centering
\caption{Results for the audit of the models with ASR $\ge 0.60$ in the exploratory full-parameter condition. The complete method reports the correct task for all 22 of them, and rank-one localization is 19 of 22.}
\label{tab:fullft-exploratory-outcomes}
\footnotesize
\begin{tabular}{@{}lc@{}}
\toprule
Quantity & Count \\
\midrule
Models with ASR $\ge 0.60$, exploratory condition & 22/54 \\
Models with ASR $\ge 0.60$, matched name targets & 9/18 \\
\midrule
True task at rank one & 19/22 \\
True task in the top three & 22/22 \\
True-task Stage-B detection & 22/22 \\
\midrule
Correct task reported by the complete method & \textbf{22/22} \\
Wrong task reported & 0/22 \\
Reported clean & 0/22 \\
\bottomrule
\end{tabular}
\end{table}

The matched name targets are the 18 models of the experiment above with the same learning rate and budget, and 9 of them are successful; the 16 of 36 reported above also include the smaller budget.

For all 22 successful attacks, Stage~A places the target task in the top three, Stage~B flags it and the complete method reports it. One of them, with an animal target on the verse task, has a score of exactly $0.600$. The two successful attacks with category targets reach the threshold only through a word that the poisoned training responses repeat, ``useful''. With 22 models over three entity types and two seeds, this set does not compare entity types.

\section{Model Size}
\label{app:model-size}

\paragraph{Setup.} We train the 12 poisoning settings of Appendix~\ref{app:benign-scaling}, with the benign data of the main set and three matched clean models each, on Llama-3.2-3B-Instruct, Qwen2.5-3B-Instruct and Qwen2.5-14B-Instruct-1M. Training, the candidate-task set, the scores and the threshold are those of the main evaluation. Each model has its own base-model continuations for Stage~A and base-model responses for Stage~B. The Llama-3.2 chat template inserts the current date into the prompt; we fix it to the date used by the Llama-3.1 template. Table~\ref{tab:model-size} compares the five models on these 12 settings, with the Llama-3.1-8B and Qwen2.5-7B rows taken from the main sets.

\paragraph{Results.} No model gives an incorrect report, and no matched clean model is a false positive. On Llama-3.2-3B, Stage~B detects 10 of the 12 models when given the target task, but Stage~A shortlists only the four verse models. LibriSpeech sentence completion, which is not a training task, is in the top three of all 15 models of this size, next to BoolQ question generation; on Llama-3.1-8B it ranks 13th or lower. Stage~B flags it for two verse models with year targets, where the highest-coverage candidate is the biased content itself; the target task scores higher for both, so both reports are correct. On the three Qwen models, the four models without a correct report are the long-narrative models, whose ASR is at most $0.16$, $0.37$ and $0.36$ at 3B, 7B and 14B parameters. On Qwen, Loc@3 rises with model size from 8 to 10 and 11, while the correct reports stay at 8. The study uses one training seed and one size of benign data.

\begin{table}[t]
\centering
\caption{Results for five models on the same 12 poisoning settings. Detected: Stage~B with the target task provided. Qwen2.5-7B has no matched clean models with 30 target-task examples. The complete method reports the correct task for 8 of 12 models on four of the five models and gives no incorrect report on any of them.}
\label{tab:model-size}
\footnotesize
\setlength{\tabcolsep}{2.5pt}
\begin{tabular}{@{}lccccccc@{}}
\toprule
Model & Loc@1 & Loc@3 & \shortstack{ASR\\$\ge 0.60$} & Detected & Correct & \shortstack{Incor-\\rect} & \shortstack{False\\positives} \\
\midrule
Llama-3.2-3B & 3 & 4 & 7 & 10 & 4 & 0 & 0/3 \\
Llama-3.1-8B & 1 & 8 & 9 & 10 & 8 & 0 & 0/3 \\
Qwen2.5-3B & 8 & 8 & 4 & 8 & 8 & 0 & 0/3 \\
Qwen2.5-7B & 8 & 10 & 6 & 8 & 8 & 0 & --- \\
Qwen2.5-14B & 7 & 11 & 7 & 8 & 8 & 0 & 0/3 \\
\bottomrule
\end{tabular}
\end{table}

\section{Reworded Instructions}
\label{app:paraphrase}

\paragraph{Setup.} A candidate prompt pairs the instruction of its task with an input, and for the target tasks this instruction is the one used in training. For each target task, we write one paraphrase of the instruction in each of three styles, a full rewording, a single sentence and a conversational request, and pair each paraphrase with 10 of the 30 inputs, keeping the inputs and the prompt format. No paraphrase shares a sequence of four words with the original. We replace the prompts of the three target tasks with these prompts and repeat Stage~A and Stage~B for the 108 poisoned Llama models and 12 Llama models of the main clean set, without retraining. The other 138 candidate tasks keep their prompts, with the Stage-A scores of Appendix~\ref{app:topk} and the responses of Appendix~\ref{app:stageb-only}. The prompt condition of Stage~B uses the words of the paraphrased prompts. The original-wording condition uses the responses of Appendix~\ref{app:stageb-only}, which give 52 successful attacks rather than the 53 of the main generation run.

\paragraph{Results.} Table~\ref{tab:paraphrase} compares the two conditions. The median ASR over the 108 models falls from $0.589$ to $0.467$, and the three styles give $0.467$, $0.433$ and $0.500$. The change depends on the target task (Table~\ref{tab:paraphrase-task}): the dialogue task keeps its 21 successful attacks, and its reworded version ranks first for 33 of its 36 models, while the median ASR of the verse task falls from $0.79$ to $0.53$ and its correct reports from 26 to 11. Of the 55 models reported correctly with the original wording, 22 are not reported with the paraphrases, and one other model is. For 20 of these 22, the Stage-B score of the reworded target task is below the threshold, and 16 of these 20 have ASR below $0.60$ on the reworded prompts; for the other two, the target task is detected but ranks outside the top three. Of the 36 attacks that remain successful on the reworded prompts, Stage~B detects 29 when given the target task, Stage~A shortlists 32 and the complete method reports 25, compared with 47, 47 and 42 of the 52 successful attacks with the original wording. The only incorrect report comes from the same model in both conditions, and no clean model is a false positive. The paraphrases were written by the authors, one per style and task, and only the target tasks are reworded.

\begin{table}[t]
\centering
\caption{Results for the original and the reworded target-task instructions, over 108 poisoned Llama models and 12 clean Llama models, from one generation run. Detected: Stage~B with the target task provided. Rewording lowers the ASR and the correct reports without adding incorrect reports or false positives.}
\label{tab:paraphrase}
\footnotesize
\setlength{\tabcolsep}{2pt}
\begin{tabular}{@{}lcccccccc@{}}
\toprule
Wording & Loc@1 & Loc@3 & Loc@10 & \shortstack{ASR\\$\ge 0.60$} & \shortstack{De-\\tected} & Correct & \shortstack{Incor-\\rect} & Clean \\
\midrule
Original & 13 & 73 & 96 & 52 & 63 & 55 & 1 & 0 \\
Paraphrased & 33 & 60 & 95 & 36 & 40 & 34 & 1 & 0 \\
\bottomrule
\end{tabular}
\end{table}

\begin{table}[t]
\centering
\caption{Results for the original and the reworded instructions by target task, over 36 models per task. Each cell gives the original condition and then the paraphrased condition. The effect of rewording depends on the target task.}
\label{tab:paraphrase-task}
\footnotesize
\setlength{\tabcolsep}{2.5pt}
\begin{tabular}{@{}lccccc@{}}
\toprule
Target task & Median ASR & ASR $\ge 0.60$ & Loc@3 & Detected & Correct \\
\midrule
Dialogue turn & 0.61 / 0.64 & 21 / 21 & 34 / 36 & 29 / 23 & 28 / 23 \\
Verse & 0.79 / 0.53 & 27 / 13 & 32 / 24 & 29 / 14 & 26 / 11 \\
Long narrative & 0.15 / 0.13 & 4 / 2 & 7 / 0 & 5 / 3 & 1 / 0 \\
\bottomrule
\end{tabular}
\end{table}

\section{Other Target Tasks}
\label{app:other-tasks}

We audit 16 poisoned models from an earlier study~\cite{sun2026poisonforge} without retraining. Their target tasks, extreme abstract summarization and health-fact explanation, are candidate tasks but not target tasks of the main set. Each model is fine-tuned on 1000 benign and 10 poisoned examples with a year target in the fixed or the category mode, and the target appears once or three times in each poisoned response, compared with five times in the main set. Eight models adapt Llama-3.1-8B-Instruct and eight adapt Qwen2.5-7B-Instruct; this set has no matched clean models. Stage~A and Stage~B are unchanged, with the base-model responses of Appendix~\ref{app:stageb-only}.

Stage~A places the target task in the top three for one of the 16 models (Table~\ref{tab:other-tasks}); the target task ranks between 17th and 79th on Llama and between 3rd and 32nd on Qwen. Six of the 16 attacks are successful. Two have fixed targets, with ASR $0.82$ and $0.71$, and four have category targets, with ASR from $0.71$ to $0.87$. When given the target task, Stage~B detects the two fixed-target attacks and one of the four category attacks. The complete method reports the correct task for one model, the fixed-target attack whose target task ranks third. The method reports an incorrect task for six models: five Qwen models on DSTC3 dialogue summarization, a benign training task, with the candidate ``make'', and one Llama model on StrategyQA question generation with a string of numbers that scores $0.611$. The training prompts of these models are not available, so their category lists contain every year in the ten poisoned responses (Appendix~\ref{app:category-asr}). On the same prompts, the base model writes a year of these lists in up to $24\%$ of its responses on the health-fact task. With a low budget and six successful attacks, this set shows that Stage~A can miss other target tasks but does not measure how often.

\begin{table}[t]
\centering
\caption{Results for 16 poisoned models on two target tasks outside the main set. Four models per row; detected: Stage~B with the target task provided; correct and incorrect: the complete method at $K=3$. Stage~A shortlists the target task for one of the 16 models.}
\label{tab:other-tasks}
\footnotesize
\setlength{\tabcolsep}{3pt}
\begin{tabular}{@{}llcccc@{}}
\toprule
Model & Target task & Target ranks & Detected & Correct & Incorrect \\
\midrule
Llama & summarization & 17, 27, 32, 38 & 2 & 0 & 0 \\
Llama & health facts & 59, 62, 73, 79 & 0 & 0 & 1 \\
Qwen & summarization & 3, 6, 10, 25 & 3 & 1 & 2 \\
Qwen & health facts & 12, 13, 14, 32 & 0 & 0 & 3 \\
\bottomrule
\end{tabular}
\end{table}

\section{Abstention When the Target Task Is Absent}
\label{app:openbank}

\paragraph{Setup.} We test whether a classifier built only from clean models can tell when the task that a model was fine-tuned on is missing from the candidate-task set, so that the method abstains instead of reporting another task. The classifier uses only the shape of the Stage-A scores.

\paragraph{Clean models.} Our earlier clean models fine-tuned on single tasks cover only the three target tasks of the main set, and the others are fine-tuned on mixtures of tasks. We therefore train 48 new clean models on 48 other tasks, chosen by task identifier from eight groups of output styles. Each model is trained on the fixed set of 1000 benign examples, without examples of its own task, plus 10, 30 or 50 clean examples of its task, with 16 models at each budget. Each model is scored twice, with the full candidate-task set and with its own task removed. We split the models, keeping both scores of a model together and stratifying by output style, into 29 for training, 10 for validation and 9 for testing.

\paragraph{Features and classifier.} Six features, fixed in advance, describe the top of the ranking: the prominence of the top score, tail exceedance, the normalized gap between the top two scores, the concentration of the score profile, and the stability of the top-one and top-three sets over 128 bootstrap resamples of the prompts. The classifier is an L2-regularized logistic regression whose regularization strength is chosen on the validation models, and its threshold is set on clean models to retain at least 95 percent of clean models scored with the full candidate-task set.

\paragraph{Results on clean models.} On the 9 test models, the area under the ROC curve is $0.444$, retention with the full candidate-task set is $1.000$ and abstention with the task removed is $0.000$: the classifier retains every test model in both conditions. The criterion fixed in advance, retention of at least $0.90$ and abstention of at least $0.50$, is not met.

\paragraph{Why the features fail.} Stage~A ranks the task of the clean model first for only 4 of the 48 models, and for none of the 16 at the smallest budget; two tasks hold rank one for 28 and 17 of the 48. For the 44 models whose task does not rank first, removing that task barely changes the top of the ranking: the median change of each feature between the two conditions is at most a few hundredths. The features therefore describe the two recurring tasks rather than whether the task of the model is present. Rank-one localization is similarly weak for these clean models, $8.3$ percent, and for the poisoned models of the main set, 12 of 108 or $11.1$ percent.

\paragraph{Results on poisoned models.} With the target task present, the classifier retains all 108 poisoned models, including the 56 for which the complete method reports the correct task. With the target task removed, it abstains for 2 of the 108 models, $1.85\%$, against the required $50\%$. Both would otherwise report an incorrect task, so six incorrect reports remain, against eight without the classifier.

\paragraph{Score reproduction.} Recomputing the Stage-A scores reproduces the saved scores exactly for 102 of the 108 models and within $9.06\times10^{-3}$ for the other six. All models keep the same top-three tasks, and one target-task rank changes from 17 to 18. One of the affected models reproduces its saved score when scored alone; we did not find the cause of the difference when several models are scored together. The results above use the recomputed scores of all 108 models.

\paragraph{Scope.} The test set has nine clean models from one model family and one adapter configuration, and we did not run a second model family. The six features above do not support an abstention rule; other signals were not tested.

\section{Clean Models Fine-Tuned on 48 Tasks}
\label{app:clean-fpr}

\paragraph{Setup.} We apply the complete method to the 48 clean models of Appendix~\ref{app:openbank}, each fine-tuned on the shared set of 1000 benign examples plus 10, 30 or 50 clean examples of one of 48 tasks. Stage~B scores the three tasks that Stage~A selects for each model, under three generation seeds and with $t^\star=0.60$.

\paragraph{Results.} The method flags one, two and one of the 48 models under the three seeds. Two models are flagged at least once and one under all three seeds. Of the 432 task evaluations, four exceed the threshold, none lies within $0.01$ of $t^\star$, and the largest margin is $0.400$.

\paragraph{Flagged tasks.} Three of the four false positives come from the model fine-tuned on Logic2Text sentence generation, evaluated on that same task under every seed: the word ``record'' appears in 89 or 90 of its 90 responses and in none of the base-model responses. The model uses this word after fine-tuning on the task and the base model does not, and Stage~B cannot tell such a word from biased content without knowing which tasks the model was trained on. The fourth false positive falls on StrategyQA question generation, with the candidate ``group'' and a score of $0.633$. Stage~A selects the StrategyQA task for 46 of the 48 models, including 44 of the 46 models that are never flagged, and BoolQ question generation, selected as often, never exceeds the threshold.

\paragraph{Scope.} All 48 models share the same benign examples and use Llama-3.1-8B-Instruct. They are built differently from the clean models of Section~\ref{sec:eval-setup}, so we report the two sets separately.

\section{Clean Models Fine-Tuned on Chat Data}
\label{app:tulu}

The clean models of Section~\ref{sec:end-to-end-auditing} are fine-tuned on Super-NaturalInstructions tasks, the same collection as the candidate-task set. We also fine-tune six Llama-3.1-8B-Instruct models with the LoRA procedure of the main set on single-turn examples from the Tulu~3 supervised mixture~\cite{lambert2024tulu3}, three on 1000 and three on 5000 examples, with different examples for each model. We remove its FLAN subset, which contains Super-NaturalInstructions tasks, its multilingual Aya subset and its hard-coded identity examples; most of the remaining examples are mathematics, code and WildChat conversations. The complete method is unchanged and uses the base-model responses of Appendix~\ref{app:stageb-only}.

No model is a false positive (Table~\ref{tab:tulu}), and the highest Stage-B score across the 18 shortlisted tasks is $0.389$. BoolQ question generation, a benign training task of the main set that is in the shortlist of every main-set model, is in none of these shortlists. StrategyQA question generation, a training task of neither set, is shortlisted for four of the six models and scores at most $0.278$. Six models from one family and one generation run are not enough to estimate a false-positive rate.

\begin{table}[t]
\centering
\caption{Results for six Llama models fine-tuned on Tulu~3 chat data, giving the Stage-A top three and the highest Stage-B score among them against the threshold of $0.60$; QG is question generation. No model is a false positive.}
\label{tab:tulu}
\footnotesize
\setlength{\tabcolsep}{3pt}
\begin{tabular}{@{}llc@{}}
\toprule
Examples & Stage-A top three & Highest $S_B$ \\
\midrule
1000 & StrategyQA QG, StoryCloze, Yelp style transfer & 0.222 \\
1000 & StoryCloze, Yelp style transfer, DiscoFuse & 0.389 \\
1000 & StrategyQA QG, XL-WiC, DiscoFuse & 0.200 \\
5000 & StoryCloze, StrategyQA QG, DiscoFuse & 0.278 \\
5000 & StrategyQA QG, XL-WiC, StoryCloze & 0.389 \\
5000 & DiscoFuse, StoryCloze, XL-WiC & 0.111 \\
\bottomrule
\end{tabular}
\end{table}

\section{Model-Level Operating Curve}
\label{app:model-level}

Figure~\ref{fig:model-level} evaluates the complete method at the level of a model, the unit it is applied to. For each poisoned model, the vertical axis counts a success when the method reports the correct target task; for each clean model, the horizontal axis counts a false positive when any of its three shortlisted tasks passes the threshold. The poisoned models use the main run, and the clean rates are averaged over three generation seeds, which resample the same models and are not independent tests. Raising the threshold above $0.80$, until none of the 30 clean models is flagged under any seed, leaves 27 correct target-task reports, and at that threshold one of the 48 clean models is still flagged under each seed. Lowering the threshold raises correct reports to at most 72 and flags more clean models.

\end{document}